# The exact solution to the Shallow water Equations Riemann problem at width jumps in rectangular channels

Giada Varra[1], Veronica Pepe[2], Luigi Cimorelli[3], Renata Della Morte[4], Luca Cozzolino[5]

## Abstract

*Riemann problems at geometric discontinuities are a classic and fascinating issue of hydraulics. In the present paper, the complete solution to the Riemann problem of the one-dimensional Shallow water Equations at monotonic width discontinuities is presented. This solution is based on the assumption that the relationship between the states immediately to the left and to the right of the discontinuity is a stationary weak solution of the one-dimensional variable-width Shallow water Equations. It is demonstrated that the solution to the Riemann problem always exists and it is unique, but there are cases where three solutions are possible. The appearance of multiple solutions is connected to a phenomenon, known as hydraulic hysteresis, observed for supercritical flow in contracting channel. The analysis of a Finite Volume numerical scheme from the literature*

[1] Ph. D. Stud., Dept. of Engrg., Parthenope Univ., Centro Direzionale di Napoli – Is. C4, 80143 Napoli, Italy. E-mail: giada.varra@uniparthenope.it
[2] Ph. D., Dept. of Engrg., Parthenope Univ., Centro Direzionale di Napoli – Is. C4, 80143 Napoli, Italy. E-mail: veronica.pepe@uniparthenope.it
[3] Sen. Res., Ph. D., Dept. of Civil, Architectural and Environmental Engrg., Federico II Univ., via Claudio 21, 80125 Napoli, Italy. E-mail: luigi.cimorelli@unina.it
[4] Full Prof., Ph. D., Dept. of Engrg., Parthenope Univ., Centro Direzionale di Napoli – Is. C4, 80143 Napoli, Italy. E-mail: renata.dellamorte@uniparthenope.it
[5] Ass. Prof., Ph. D., Dept. of Engrg., Parthenope Univ., Centro Direzionale di Napoli – Is. C4, 80143 Napoli, Italy. E-mail: luca.cozzolino@uniparthenope.it

*(Cozzolino et al. 2018b) shows that the algorithm captures the solution with supercritical flow through the width discontinuity when multiple solutions are possible. Interestingly, the one-dimensional variable-width Shallow water Equations are formally identical to the one-dimensional Porous Shallow water Equations, implying that the exact solutions and the numerical scheme discussed in the present paper are relevant for two-dimensional Porous Shallow water numerical models aiming at urban flooding simulations. The exact solution presented here may be used not only as a benchmark, but also as a guide for the construction of new algorithms, and it can be even embedded in an exact solver.*

**Subject headings**: Shallow water Equations; Porous Shallow water Equations; Riemann problem; Width; Channel contraction; Channel expansion; Hydraulic hysteresis.

Corresponding author: Luca Cozzolino

E-mail address: luca.cozzolino@uniparthenope.it

Address: Dipartimento di Ingegneria, Università degli Studi di Napoli Parthenope, Isola C4, 80143 Napoli (Italy)

Phone: +390815476723.

To the beloved memory of Mena and Tonino.

# 1. Introduction

The Saint Venant Equations are a mathematical model widely used for the simulation of flood wave propagation in natural and artificial channels (Cunge et al. 1980). One of the main model assumptions is that the cross-section shape continuously varies along the channel, and this allows an easy study of the mathematical model in the context of balance hyperbolic systems of equations with continuously varying source terms. The solution to the corresponding Riemann problem, i.e. the solution of the initial value problem with discontinuous initial flow conditions, is of great practical relevance. For instance, the dam-break problem (Stoker 1957), which is the special Riemann problem where water is initially at rest, may serve to model flood propagation after a dam collapse. In addition, the Riemann problem is the building block for the construction of numerical models that solve the Saint Venant Equations under the assumption of continuously varying cross-section (García-Navarro and Vázquez-Cendón 2000, Goutal and Maurel 2001, García-Navarro et al. 2002, Lee and Wright 2010, Cozzolino et al. 2012, Xing 2016, e.g.).

Unfortunately, real-world channels exhibit non-prismatic reaches where the cross-section shape and dimensions rapidly vary, due to the presence of structures such as bridge piles, Venturi flumes, sluice gates, culverts, weirs, check-dams, contractions, and expansions (Chow 1959). When geometric discontinuities are present, the Saint Venant equations can be solved by separately considering the channel reaches to the left and to the right of the geometric discontinuity, and by connecting the two reaches through appropriate internal boundary conditions (Cunge et al. 1980). If the flow variables to the left and to the right of the geometric transition are far from the equilibrium, it is possible to consider a Riemann problem where both flow and geometric characteristics have a jump, and the corresponding study can be accomplished in the context of non-conservative systems of hyperbolic equations (Dal Maso et al. 1995). The Riemann problem at geometric discontinuities also has a great practical relevance because it corresponds to the transient generated by rapid manoeuvres at sluice-gates or by the impact of a bore against a hydraulic structure. Usually, this

Riemann problem is studied considering the one-dimensional Shallow water Equations, which are the restriction of the Saint Venant Equations to the case of rectangular cross section. In this context, many different types of geometric discontinuities have been considered in the literature, such as sluice gates (De Marchi 1945, Montuori and Greco 1973, Cozzolino et al. 2015), bed steps (Alcrudo and Benkhaldoun 2001, LeFloch and Than 2011, and Han and Warnecke 2014), free or submerged weirs (Guerra et al. 2008), and obstructions or constrictions (Cozzolino et al. 2017, Pepe et al. 2019).

Among the other geometric discontinuities, abrupt width jumps in rectangular channels have been frequently studied. Ostapenko (2012a) solved the dam-break problem at contracting channels under the assumption of head and discharge invariance through the contraction. The same Author (Ostapenko 2012b) generalized his results considering both contracting and expanding channels, finding that the problem is ill-posed for some combination of left and right flow depths when the assumption of head and discharge invariance through the geometric discontinuity is used. Kovyrkina and Ostapenko (2013) considered the dam-break at contracting channels with a prescribed head loss. Their solutions compared well with laboratory experiments in Degtyarev et al. (2014). Cozzolino et al. (2018b) dropped the assumption of head and discharge invariance through the width discontinuity, and assumed that the relationship between the states immediately to the left and to the right of the width discontinuity was a stationary weak solution of Shallow water Equations in variable-width rectangular channels. This assumption allowed introducing head loss through the width discontinuities by means of the hydraulic jump mechanism. The Authors supplied the exact solution for this dam-break problem, showing that the solution existed and was unique for each initial condition. Successively, Cozzolino et al. (2018a) considered a special Riemann problem at channel rapid variations with right dry bed state. The Authors found that the problem may exhibit multiple solutions if the flow impinging a rapid contraction is supercritical, and connected this finding with the hydraulic hysteresis phenomenon (Defina and Susin 2006). Valiani and Caleffi (2019) replicated the dam-break solution by Cozzolino et al. (2018b), giving additional mathematical details that were implicit in Cozzolino et al. (2018b). Finally, Goudiaby and Kreiss (2020) considered a class of Riemann problems at sudden expansions with Borda-Carnot losses and subcritical flow, demonstrating that there are initial conditions for which this problem has no solution.

Despite the interesting results from the literature, it is evident that existing works on the Riemann problem at width discontinuities are focused on restrictive classes of problems (the dam-break, the Riemann problem with right dry bed, and the Riemann problem with subcritical flow and Borda-Carnot losses) and that there is need for a more general discussion. For this reason, the complete solution to the Riemann problem of the one-dimensional variable-width Shallow water Equations is given in the present paper, under the assumption that the relationship between the states immediately to the left and to the right of the width discontinuity is a stationary weak solution of the Shallow water Equations. It is demonstrated that the solution to the Riemann problem always exists, but that there are certain initial conditions for which multiple solutions are possible. In addition, it is shown that the solution multiplicity is connected to the hydraulic hysteresis in contracting channels (Defina and Viero 2010, Viero and Defina 2017). Finally, the numerical method by Cozzolino et al. (2018b) is analysed, showing that the structure of the algorithm forces the numerical solution with supercritical flow through the width discontinuity when multiple solutions are possible.

Interestingly, the relevance of the results presented here goes well beyond the simulation of flood wave propagation in channels, because the one-dimensional variable-width Shallow water Equations are formally identical to the one-dimensional Porous Shallow water Equations (Guinot and Soares-Frazão 2006, Sanders et al. 2008). The porosity Riemann problem solution is the building block for the construction of two-dimensional Porous Shallow water Equations numerical schemes (Guinot and Soares-Frazão 2006, Finaud-Guyot et al. 2010, Cea and Vázquez-Cendón 2020, and Ferrari et al. 2017, e.g.), and this implies that the exact solutions and the numerical scheme discussed in the present paper may be directly incorporated into current two-dimensional numerical models for urban flooding simulations.

The work is structured as follows. In Section 2, the one-dimensional Shallow water Equations in variable-width rectangular channels are presented, and the inner boundary condition through width jumps is commented. In Section 3, the Riemann problem at width jumps is defined, the complete solution of the Riemann problem is presented, and a result about existence and multiplicity is proven. In Section 4, the numerical algorithm by Cozzolino et al. (2018b) is applied to numerous Riemann problems. In Section 5, the connection between Riemann solution multiplicity and the hydraulic hysteresis, the disambiguation of Riemann multiple solutions, and the flaws of the aforementioned algorithm, are discussed. In Section 6, final conclusions are drawn.

## 2. Mathematical model

Under the assumptions of horizontal bed and negligible friction, the one-dimensional Shallow water Equations in variable-width rectangular channels can be written as

$$(1) \quad \frac{\partial B\mathbf{u}}{\partial t} + \frac{\partial B\mathbf{f}(\mathbf{u})}{\partial x} + \mathbf{h}(\mathbf{u})\frac{\partial B}{\partial x} = 0.$$

In Eq. (1), the meaning of the symbols is as follows: $x$ is the longitudinal coordinate; $t$ is the time variable; $B(x)$ is the local channel width; $\mathbf{u}(x,t) = \begin{pmatrix} h & hu \end{pmatrix}^T$ is the vector of the unit-width conserved variables, where $h(x, t)$ is the flow depth, $u(x, t)$ is the cross-section averaged velocity, and $T$ is the matrix transpose symbol; $\mathbf{f}(\mathbf{u}) = \begin{pmatrix} hu & 0.5gh^2 + hu^2 \end{pmatrix}^T$ is the vector of the one-dimensional Shallow water Equations fluxes, where $g$ = 9.81 m/s² is the gravitational acceleration;

finally, $\mathbf{h}(\mathbf{u}) = \begin{pmatrix} 0 & -0.5gh^2 \end{pmatrix}^T$ is a vector that takes into account the forces exerted on the flow by the channel walls.

The Riemann problem is the initial value problem where Eq. (1) is solved with discontinuous initial conditions

$$(2) \quad \mathbf{u}(x,0) = \begin{cases} \mathbf{u}_L, & x < 0 \\ \mathbf{u}_R, & x > 0 \end{cases}$$

and discontinuous width

$$(3) \quad B(x) = \begin{cases} B_L, & x < 0 \\ B_R, & x > 0 \end{cases}.$$

In Eqs. (2) and (3), $\mathbf{u}_L = \begin{pmatrix} h_L & h_L u_L \end{pmatrix}^T$ and $B_L$ are the flow initial conditions and the channel width to the left of the geometric discontinuity in $x = 0$, respectively, while $\mathbf{u}_R = \begin{pmatrix} h_R & h_R u_R \end{pmatrix}^T$ and $B_R$ are the flow initial conditions and the channel width to the right of $x = 0$.

The solution $\mathbf{u}(x,t)$ of the Riemann problem expressed by Eqs. (1)-(3) is self-similar, and this implies that it exists a vector function $\mathbf{w}(\bullet)$ of a real variable such that $\mathbf{u}(x,t) = \mathbf{w}(x/t)$ for $t > 0$. It follows that the Riemann solution consists of a sequence of constant states separated by moving or standing waves (Cozzolino et al. 2018b, Pepe et al. 2019). The condition of Eq. (3) implies that Eq. (1) reduces to the classic one-dimensional Shallow water Equations

$$(4) \quad \frac{\partial \mathbf{u}}{\partial t} + \frac{\partial \mathbf{f}(\mathbf{u})}{\partial x} = 0$$

for $x < 0$ and $x > 0$ in the Riemann problem. For this reason, the elementary waves moving to the left or to the right of the initial discontinuity in $x = 0$ coincide with the shocks and the rarefactions of the classic Shallow water Equations (see the corresponding discussion in LeFloch and Thanh 2007 for the case of the Riemann problem at the bed step). In the $(h, u)$ plane, the symbols $S_i(\mathbf{u}_{ref})$ and $S_i^B(\mathbf{u}_{ref})$ denote the curves of the states $\mathbf{u}$ connected to the reference state $\mathbf{u}_{ref} = \begin{pmatrix} h_{ref} & h_{ref} u_{ref} \end{pmatrix}^T$ by the direct and backward shocks contained into the $i$-th (with $i = 1, 2$) characteristic field of the classic one-dimensional Shallow water Equations, respectively, while the symbols $R_i(\mathbf{u}_{ref})$ and $R_i^B(\mathbf{u}_{ref})$ denote the curves of the direct and backward rarefactions. The corresponding equations in the $(h, u)$ plane are resumed in Appendix A. The direct and backward $i$-waves are defined as $T_i(\mathbf{u}_{ref}) = R_i(\mathbf{u}_{ref}) \cup S_i(\mathbf{u}_{ref})$ and $T_i^B(\mathbf{u}_{ref}) = R_i^B(\mathbf{u}_{ref}) \cup S_i^B(\mathbf{u}_{ref})$, respectively.

In addition to the elementary curves of the Shallow water Equations, it is convenient to define the constant discharge curve

(5) $CD(\mathbf{u}_{ref}): \quad u = (h_{ref} u_{ref})/h$

as the locus of the $(h, u)$ plane consisting of the states $\mathbf{u}$ that have the same unit-width discharge of $\mathbf{u}_{ref}$.

When the reference state $\mathbf{u}_{ref}$ is supercritical, a hash symbol superscript denotes the subcritical state $\mathbf{u}_{ref}^{\#}$ that is conjugate to $\mathbf{u}_{ref}$ by means of the hydraulic jump condition. It is evident that $\mathbf{u}_{ref}^{\#}$ is at the intersection, distinct from $\mathbf{u}_{ref}$, of the $CD(\mathbf{u}_{ref})$ and $S_1(\mathbf{u}_{ref})$ curves when $u_{ref} > 0$, or at the intersection between $CD(\mathbf{u}_{ref})$ and $S_2^B(\mathbf{u}_{ref})$ when $u_{ref} < 0$.

## 2.1 Standing wave at width jumps

The geometric discontinuity in $x = 0$ is a mathematical conceptualization used to approximate a rapid width variation that is much shorter than the channel. Despite its infinitesimal length, the geometric discontinuity considered in the present paper has an internal structure consisting of a continuous smooth monotonic variation of the channel width from $B(0^-) = B_L$ to $B(0^+) = B_R$ (Figure 1). Only the case $B_L < B_R$ is considered here because the solutions corresponding to $B_L > B_R$ can be obtained by simply mirroring the procedure outlined in the following, while the case $B_L = B_R$ is trivial because it reduces to the Riemann problem of the classic Shallow water Equations (Stoker 1957, Toro 2001).

The presence of a geometric rapid variation induces an additional flow discontinuity with respect to the classic moving rarefactions and shocks. This standing wave (SW) establishes a relationship between the states $\mathbf{u}_1 = (h_1 \quad h_1 u_1)^T = \mathbf{u}(0^-, t)$ and $\mathbf{u}_2 = (h_2 \quad h_2 u_2)^T = \mathbf{u}(0^+, t)$ immediately to the left and to the right of $x = 0$, respectively. The SW relationship is stationary because the self-similarity of the Riemann problem solution implies that $\mathbf{u}_1 = \mathbf{w}(0^-)$ and $\mathbf{u}_2 = \mathbf{w}(0^+)$ are constant in time. For this reason, stationary solutions of Eq. (1) have been proposed in the literature (Castro et al. 2007, Cozzolino et al. 2018b, Pepe et al. 2019) as good candidates to supply a relationship between $\mathbf{u}_1$ and $\mathbf{u}_2$, and they will be used in the present work. These stationary solutions are characterized by the invariance of the discharge $Q = Bhu$, implying that this condition must be always assumed for the definition of the SW.

When the flow through the discontinuity is smooth, the specific energy $E = h + u^2/(2g)$ is also invariant through the rapid width variation. For a given reference state $\mathbf{u}_{ref}$ immediately to the left [right] of the geometric discontinuity, a nought superscript denotes the state $\mathbf{u}_{ref}^0$ immediately to the right [left] of the discontinuity that is conjugate to $\mathbf{u}_{ref}$ by the conditions of discharge and energy invariance. It is evident that both $\mathbf{u}_{ref}$ and $\mathbf{u}_{ref}^0$ are either supercritical or subcritical. Of course, energy is dissipated through the width discontinuity when a hydraulic jump is present, and only the invariance of the discharge through the geometric discontinuity is preserved in this case.

Note that the additional standing wave at width jumps is connected to the presence of the non-conservative product $\mathbf{h}(\mathbf{u})\partial B/\partial x$. The solution of systems of hyperbolic differential equation with non-conservative products has been considered in Dal Maso et al. (1995), where the need of defining a generalized Rankine-Hugoniot condition through the jumps is underlined. The definition of the standing wave at width jumps proposed here is actually equivalent to a generalized Rankine-Hugoniot condition.

[Insert Figure 1 about here]

*2.1.1 Froude limits at the contraction*

Under the assumption of energy invariance through the geometric discontinuity, the Yarnell's criterion (Yarnell 1934) states that the smooth passage of the flow with Froude number $F = u/\sqrt{gh}$ from the section of width $B$ to the section of width $B_0$ (with aspect ratio $B_0/B < 1$) implies that $0 \leq |F| \leq K_{sb}(B_0/B) < 1$ or $|F| \geq K_{sp}(B_0/B) > 1$. The subcritical Froude limit $K_{sb}(B_0/B)$ and the supercritical Froude limit $K_{sp}(B_0/B)$ depend on the aspect ratio $B_0/B < 1$ of the contraction, and the corresponding expressions are reported in Appendix B. Of course, the Yarnell's criterion has no meaning when $B_0/B \geq 1$, because the flow has always sufficient energy to pass through expansions.

In addition to the Froude limits $K_{sb}(B_0/B)$ and $K_{sp}(B_0/B)$, it is useful to consider the conjugate limits $K_{jump}(B_0/B)$ and $K_{sp}^{\#}(B_0/B)$. The conjugate limit $K_{jump}(B_0/B)$ is defined as the Froude number of the supercritical flow connected by a hydraulic jump to the subcritical flow with Froude number $K_{sb}(B_0/B)$. On the other hand, the conjugate limit $K_{sp}^{\#}(B_0/B)$ is defined as the Froude number of the subcritical flow connected by a standing hydraulic jump to the supercritical flow with Froude number $K_{sp}(B_0/B)$. The expressions of the conjugate Froude limits $K_{jump}(B_0/B)$ and $K_{sp}^{\#}(B_0/B)$ are reported in Appendix B, as well. A simple tabulation demonstrates that $K_{sb}(B_0/B) < K_{sp}^{\#}(B_0/B) < 1 < K_{sp}(B_0/B) < K_{jump}(B_0/B)$ is always satisfied.

Associated to the Froude limits and to the aspect ratio $B_0/B < 1$, it is possible to consider the following curves of the (*h*, *u*) plane

$$
\begin{aligned}
&C^+: \ u = \sqrt{gh}, \quad C^-: \ u = -\sqrt{gh} \\
(6) \quad &C_{sb}^+(B_0/B): \ u = K_{sb}(B_0/B)\sqrt{gh}, \quad C_{sb}^-(B_0/B): \ u = -K_{sb}(B_0/B)\sqrt{gh}, \\
&C_{sp}^+(B_0/B): \ u = K_{sp}(B_0/B)\sqrt{gh}, \quad C_{sp}^-(B_0/B): \ u = -K_{sp}(B_0/B)\sqrt{gh}
\end{aligned}
$$

which are involved in the construction of the Riemann problem solutions.

*2.1.2 Flow configurations of the standing wave*

Six flow configurations for the standing wave through the geometric discontinuity, named from SWa to SWf, are possible under the assumption $B_L < B_R$. The flow configurations SWa, SWb, and SWc, where the flux is from left to right ($u_1 \geq 0$, $u_2 \geq 0$), are characterized as follows:

*SWa*. The flow smoothly decelerates from left to right through an expansion (see a sketch of the free surface profile in Figure 2a). The left and right states $\mathbf{u}_1$ and $\mathbf{u}_2$ are subcritical ($0 \leq F_1 \leq 1$ and $0 \leq F_2 \leq K_{sb}(B_L/B_R)$, where $F_1$ and $F_2$ are the Froude numbers associated to $\mathbf{u}_1$ and $\mathbf{u}_2$, respectively), with $F_2 \leq F_1$. Obviously, $\mathbf{u}_2 = \mathbf{u}_1^0$ and $\mathbf{u}_1 = \mathbf{u}_2^0$.

*SWb* (Figure 2b). The left state $\mathbf{u}_1$ is critical or supercritical ($F_1 \geq 1$), while the right state $\mathbf{u}_2$ is subcritical ($0 < F_2 < 1$). Through the geometric discontinuity, the supercritical accelerating flow is reverted into a subcritical decelerating flow by a stable hydraulic jump (Akers and Bokhove 2008). When $\mathbf{u}_1$ is critical, $K_{sb}(B_L/B_R) < F_2 < K_{sp}^{\#}(B_L/B_R)$.

*SWc*. The flow smoothly accelerates from left to right (see Figure 2c), and $F_1 < F_2$. The left state $\mathbf{u}_1$ is supercritical or critical ($F_1 \geq 1$), while the right state $\mathbf{u}_2$ is supercritical with $F_2 \geq K_{sp}(B_L/B_R)$. In particular, $F_2 = K_{sp}(B_L/B_R)$ when $\mathbf{u}_1$ is critical. Also in this case, $\mathbf{u}_2 = \mathbf{u}_1^0$ and $\mathbf{u}_1 = \mathbf{u}_2^0$.

The flow configurations SWd, SWe, and SWf, where the flux is from left to right ($u_1 < 0$, $u_2 < 0$), are characterized as follows:

*SWd*. The flow smoothly accelerates from right to left through a contraction (see Figure 2d), and $F_1 < F_2$ for $F_2 \neq 0$. The left state $\mathbf{u}_1$ is subcritical or critical ($-1 \leq F_1 \leq 0$), while the right state is subcritical with $-K_{sb}(B_L/B_R) \leq F_2 \leq 0$. In particular, $F_2 = -K_{sb}(B_L/B_R)$ when $\mathbf{u}_1$ is critical. It is evident that $\mathbf{u}_2 = \mathbf{u}_1^0$ and $\mathbf{u}_1 = \mathbf{u}_2^0$.

*SWe*. The flow smoothly decelerates through the rapid width variation from right to left (see Figure 2e). The left and right states $\mathbf{u}_1$ and $\mathbf{u}_2$ are supercritical with $F_2 \leq -K_{sp}(B_L/B_R)$ and $F_2 < F_1 \leq -1$. Of course, $\mathbf{u}_2 = \mathbf{u}_1^0$ and $\mathbf{u}_1 = \mathbf{u}_2^0$.

*SWf*. The flow is from right to left (see Figure 2f). The left state $\mathbf{u}_1$ is critical or subcritical ($-1 \leq F_1 < 0$), while the right state $\mathbf{u}_2$ is supercritical with $F_2 < -K_{sp}(B_L/B_R)$. The passage from supercritical to subcritical flow is forced by an unstable hydraulic jump (Akers and Bokhove 2008).

**Remark 1**. Note that these flow configurations do not correspond to those discussed in Pepe et al. (2019) because the rapid width variation considered in the present work is monotonic. This implies that a critical state can be exhibited only at one of the geometric discontinuity ends.

**Remark 2**. Alternative to the standing wave definition used here, it is possible to consider definitions where a given amount of specific energy loss through the geometric discontinuity is prescribed (Kovyrkina and Ostapenko 2013, Varra et al. 2020, Goudiaby and Kreiss 2020) or it is deduced from laboratory experiments (Formica 1955, Hager 2010). Of course, SW definitions where the specific energy may increase through the geometric discontinuity (like those considered in Guinot and Soares-Frazão 2006, Mohamed 2014, Ion et al. 2016, Ferrari et al. 2017) must be discarded (see the discussion in Cozzolino et al. 2018b).

[Insert Figure 2 about here]

## 3. Exact Riemann problem solution

In the present section, the solutions to the Shallow water Equations Riemann problem at contractions and expansions are classified, the L-M and R-M curves are defined depending on the Froude numbers of the states $\mathbf{u}_L$ and $\mathbf{u}_R$, and some examples of Riemann problem solution are presented. Finally, it is demonstrated that the solution to the Shallow water Equations Riemann problem at width jumps always exists. This solution may be unique or triple, depending on $\mathbf{u}_L$ and $\mathbf{u}_R$.

### 3.1 Structure of the Riemann problem solution

In the classic Shallow water Equations, the solution to the Riemann problem consists of three states ($\mathbf{u}_L$, $\mathbf{u}_{mid}$, and $\mathbf{u}_R$, ordered from left to right), where $\mathbf{u}_L$ and $\mathbf{u}_{mid}$ are connected by an elementary wave contained into the first characteristic field, while the states $\mathbf{u}_R$ and $\mathbf{u}_{mid}$ are connected by a wave contained into the second characteristic field (Alcrudo and Benkhaldoun 2001, Toro 2001). The structure of this solution is sketched in row 1 of Table 1, where the symbol $\mathbf{u}_i \xrightarrow{W} \mathbf{u}_j$ indicates that the left state $\mathbf{u}_i$ is connected to the right state $\mathbf{u}_j$ by means of the generic wave $W$. The indications $T_1$ and $T_2$ in the solution sketch mean that the corresponding waves may be either shocks ($S_i$ with $i = 1, 2$) or rarefactions ($R_i$ with $i = 1,2$). The classic Shallow water Equations Riemann problem can be easily solved by finding the state $\mathbf{u}_M$ at the intersection of the curves $T_1(\mathbf{u}_L)$ and $T_2^B(\mathbf{u}_R)$ in the ($h$, $u$) plane, and taking $\mathbf{u}_{mid} = \mathbf{u}_M$. If there is no intersection, the states $\mathbf{u}_L$ and $\mathbf{u}_R$ are connected to the dry bed by means of two rarefactions, respectively (Toro 2001). This case is represented in row 2 of Table 1, where **I** is the dry bed state connected to $\mathbf{u}_L$ by a rarefaction contained into the first characteristic field while **J** is the dry bed state connected to $\mathbf{u}_R$ by a rarefaction contained into the second characteristic field.

The presence of the geometric discontinuity leads to an increased number of waves and intermediate states between $\mathbf{u}_L$ and $\mathbf{u}_R$. Nonetheless, it is possible to show by enumeration that all the solutions to the Riemann problem of Eq. (1)-(3) with $B_L < B_R$ fall in one of eight possible classes, whose structure is summarised in Table 1 (rows from 3 to 10). The names of the solution classes are codified in the form SC$X_{(0)}\pm$, where the meaning is as follows: $X$ indicates the number of waves contained into the class solution; the sign plus indicates that the flow through the geometric discontinuity is directed from left to right, while the contrary is true for the sign minus; finally, the subscript 0, if present, indicates that $\mathbf{u}_1 = \mathbf{u}_L$ when the sign is plus, while $\mathbf{u}_2 = \mathbf{u}_R$ when the sign is minus. The solution classes are described in the following.

SC4+. In this class of solution configurations, which is sketched in row 3 of Table 1, the flow through the geometric discontinuity is directed from left to right. The solutions consist of five states ($\mathbf{u}_L$, $\mathbf{u}_1$, $\mathbf{u}_2$, $\mathbf{u}_{mid}$, and $\mathbf{u}_R$, ordered from left to right), which are connected by four waves ($R_1$, SWc, $T_1$, and $T_2$, from left to right). The state $\mathbf{u}_1$ is critical while the state $\mathbf{u}_2$ is supercritical with Froude number $F_2 = K_{sp}(B_L/B_R)$.

SC3$_0$+ (row 4 of Table 1). The Riemann solutions consist of four states ($\mathbf{u}_L$, $\mathbf{u}_2$, $\mathbf{u}_{mid}$, and $\mathbf{u}_R$), which are connected by three waves (SWc, $T_1$, and $T_2$). The supercritical state $\mathbf{u}_1$ coincides with $\mathbf{u}_L$, while the supercritical state $\mathbf{u}_2$ is characterized by $F_2 > K_{sp}(B_L/B_R)$.

SC3+ (row 5 of Table 1). The solutions of this class consist of four states ($\mathbf{u}_L$, $\mathbf{u}_1$, $\mathbf{u}_2$, and $\mathbf{u}_R$) connected by three waves ($T_1$, SW, and $T_2$). When SW = SWa, the states $\mathbf{u}_1$ and $\mathbf{u}_2$ are subcritical with $0 \leq F_1 < 1$ and $0 \leq F_2 < K_{sb}(B_L/B_R)$, with $F_2 < F_1$. When SW = SWb, the state $\mathbf{u}_1$ is critical ($F_1 = 1$), while $\mathbf{u}_2$ is subcritical with $K_{sb}(B_L/B_R) < F_2 < K_{sp}^{\#}(B_L/B_R)$.

SC2$_0$+ (row 6 of Table 1). The solutions of this class consist of three states ($\mathbf{u}_L$, $\mathbf{u}_2$, and $\mathbf{u}_R$) which are connected by two waves (SWb and $T_2$). The state $\mathbf{u}_1 = \mathbf{u}_L$ is supercritical ($F_1 > 1$), while $\mathbf{u}_2$ is subcritical with $0 < F_2 < 1$.

SC4- (row 7 of Table 1). In this class of solution configurations, the flow through the geometric discontinuity is directed from left to right. The Riemann solutions consist of five states ($\mathbf{u}_L$, $\mathbf{u}_{mid}$, $\mathbf{u}_1$, $\mathbf{u}_2$, and $\mathbf{u}_R$) which are connected by four waves ($T_1$, $R_2$, SWd, and $T_2$). The state $\mathbf{u}_1$ is critical ($F_1 = -1$) while the state $\mathbf{u}_2$ is subcritical with $F_2 = -K_{sb}(B_L/B_R)$.

SC3$_0$- (row 8 of Table 1). The Riemann solutions, which are characterised by $\mathbf{u}_2 = \mathbf{u}_R$, consist of four states ($\mathbf{u}_L$, $\mathbf{u}_{mid}$, $\mathbf{u}_1$, and $\mathbf{u}_R$) connected by three waves ($T_1$, $T_2$, and SW). When SW = SWe, the states $\mathbf{u}_1$ and $\mathbf{u}_2$ are supercritical with $F_1 < -1$ and $F_2 < -K_{sp}(B_L/B_R)$. When SW = SWf, the state $\mathbf{u}_1$ is critical ($F_1 = -1$), and $T_2 = R_2$.

SC3- (row 9 of Table 1). The solutions of this class consist of four states ($\mathbf{u}_L$, $\mathbf{u}_1$, $\mathbf{u}_2$, and $\mathbf{u}_R$) connected by three waves ($T_1$, SWd, and $T_2$). The states $\mathbf{u}_1$ and $\mathbf{u}_2$ are subcritical with $-1 \leq F_1 < 0$ and $-K_{sb}(B_L/B_R) \leq F_2 < 0$.

SC2$_0$- (row 10 of Table 1). The solutions of this class consist of three states ($\mathbf{u}_L$, $\mathbf{u}_1$, and $\mathbf{u}_R$) which are connected by two waves ($T_1$ and SWf). The state $\mathbf{u}_2 = \mathbf{u}_R$ is supercritical with $F_2 < -K_{sp}(B_L/B_R)$, while the state $\mathbf{u}_1$ is subcritical with $-1 < F_1 < 0$.

**Remark 3**. Note that these solution classes differ from those discussed in Pepe et al. (2019) because in the present case the internal structure of the geometric discontinuity is monotonic.

[Insert Table 1 about here]

## 3.2 L-M and R-M curves

The presence of the geometric discontinuity complicates the practical solution of the Shallow water Equations Riemann problem at width discontinuities, because the curves $T_1(\mathbf{u}_L)$ and $T_2^B(\mathbf{u}_R)$ cannot be directly used to find the Riemann problem solution. Nonetheless, it is possible to introduce the L-M and R-M curves (Marchesin and Paes-Leme 1986, LeFloch and Than 2011, Han and Warnecke 2014, Pepe et al. 2019), that generalize the curves $T_1(\mathbf{u}_L)$ and $T_2^B(\mathbf{u}_R)$.

The L-M curve consists of

- the states $\mathbf{u}$ coinciding with the states $\mathbf{u}_2$ and $\mathbf{u}_{mid}$ (with $u \geq 0$) that can be connected to $\mathbf{u}_L$ by an admissible sequence of waves (including SW with $u_1 \geq 0$ and $u_2 \geq 0$);
- the states $\mathbf{u}$ on the $T_1(\mathbf{u}_L)$ curve characterized by $u < 0$.

The R-M curve consists of

- the states $\mathbf{u}$ on the $T_2^B(\mathbf{u}_R)$ curve characterized by $u \geq 0$;
- the states $\mathbf{u}$ coinciding with the states $\mathbf{u}_1$ and $\mathbf{u}_{mid}$ (with $u < 0$) that can be connected to $\mathbf{u}_R$ by an admissible sequence of waves (including SW with $u_1 < 0$ and $u_2 < 0$).

The Riemann problem is solved by finding the intersection $\mathbf{u}_M = \begin{pmatrix} h_M & h_M u_M \end{pmatrix}^T$ between the L-M and R-M curves. The position of the intersection $\mathbf{u}_M$ along L-M and R-M determines the solution structure, namely the direction of the flow through the geometric discontinuity ($u_1 \geq 0$ and $u_2 \geq 0$ if $u_M \geq 0$, while $u_1 < 0$ and $u_2 < 0$ when $u_M < 0$), the number of waves, and the intermediate states between $\mathbf{u}_L$ and $\mathbf{u}_R$. The construction ensures that the standing wave effects are taken into account by the L-M curve when $u_M \geq 0$, while the standing wave is on the R-M curve when $u_M < 0$ (Han and Warnecke 2014).

In the following, all the example curves are referred to the aspect ratio $B_L/B_R = 0.60$, but the procedure for the curve construction is totally general. The corresponding values of the Froude limits are $K_{sb}(0.60) = 0.36$, $K_{sp}(0.60) = 2.32$, $K_{jump}(0.60) = 3.67$, and $K_{sp}^{\#}(0.60) = 0.49$.

*3.2.1 L-M curves*

Three types of L-M curves ($I_L$, $II_L$, and $III_L$) are possible, depending on the value of the Froude number $F_L = u_L/\sqrt{gh_L}$ associated to the state $\mathbf{u}_L$ (see Table 2, rows 1-3).

In the case $F_L < -2$ ($I_L$ curve), there is no state $\mathbf{u} \in T_1(\mathbf{u}_L)$ such that $u \geq 0$. This implies that $\mathbf{u}_L$ cannot be connected to the width discontinuity to its right by means of an elementary wave. Conversely, the condition $F_L \geq -2$ ($II_L$ and $III_L$ curves) implies that $\mathbf{u}_L$ may be directly connected to a state $\mathbf{u}_1$ to the left of the width discontinuity by means of $T_1(\mathbf{u}_L)$. Obviously, the condition $F_L \geq -2$ allows considering the special state $\mathbf{K}_L = \begin{pmatrix} h_{K_L} & 0 \end{pmatrix}^T \in T_1(\mathbf{u}_L)$ that is characterized by null velocity.

The construction of the L-M curves is described in the following paragraphs.

[Insert Table 2 about here]

*$I_L$ curve*. For $F_L < -2$, the L-M curve coincides with $T_1(\mathbf{u}_L)$, which lies entirely under the axis $u = 0$ in the $(h, u)$ plane. An example $I_L$ curve is plotted in Figure 3a for the state $\mathbf{u}_L$ of Row 1 in Table 3. It is evident that the $I_L$ curve is continuous and strictly decreasing in the plane $(h, u)$, and that its supremum $\mathbf{I} = \begin{pmatrix} 0 & u_L + 2\sqrt{gh_L} \end{pmatrix}^T$ coincides with the dry bed state on $R_1(\mathbf{u}_L)$,

[Insert Figure 3 about here]
[Insert Table 3 about here]

***II$_L$ curve***. In the case II$_L$, which is characterized by $F_L \in [-2, 1]$, the state **u**$_L$ lies between the curves $C^+$ and $u = -2\sqrt{gh}$ of the ($h$, $u$) plane (Figure 3b,c). This implies that it exists the critical state $\mathbf{\gamma} = \begin{pmatrix} h_\gamma & h_\gamma u_\gamma \end{pmatrix}^T$ defined as the intersection between $R_1(\mathbf{u}_L)$ and $C^+$, and that it is possible to define the following special **u**$_2$ states:

- the supercritical state $\mathbf{\gamma}_{sp} = \begin{pmatrix} h_{\gamma_{sp}} & h_{\gamma_{sp}} u_{\gamma_{sp}} \end{pmatrix}^T$ on the curve $C_{sp}^+(B_L/B_R)$, connected to $\mathbf{u}_1 = \mathbf{\gamma}$ by means of the standing wave SWc;

- the subcritical state $\mathbf{\gamma}_{sb} = \begin{pmatrix} h_{\gamma_{sb}} & h_{\gamma_{sb}} u_{\gamma_{sb}} \end{pmatrix}^T$ on the curve $C_{sb}^+(B_L/B_R)$, connected to $\mathbf{u}_1 = \mathbf{\gamma}$ by means of the standing wave SWa;

- the subcritical state $\mathbf{\gamma}_{sp}^\# = \begin{pmatrix} h_{\gamma_{sp}^\#} & h_{\gamma_{sp}^\#} u_{\gamma_{sp}^\#} \end{pmatrix}^T$, connected to $\mathbf{\gamma}_{sp}$ by means of the hydraulic jump.

From the preceding definitions, it is evident that the states **γ**$_{sb}$, and $\mathbf{\gamma}_{sp}^\#$ lie on the curve CD(**γ**$_{sp}$), while the arc of the curve CD(**γ**$_{sp}$) between **γ**$_{sb}$ and $\mathbf{\gamma}_{sp}^\#$ represents the locus of the subcritical states **u**$_2$ that can be connected to the critical state $\mathbf{u}_1 = \mathbf{\gamma}$ by means of a SWb wave. Finally, it is possible to consider the locus LS(**u**$_L$) of the subcritical states $\mathbf{u}_2 = \mathbf{u}_1^0$ that are connected by means of the standing wave SWa to the subcritical states $\mathbf{u}_1 \in T_1(\mathbf{u}_L)$ between **γ** and **K**$_L$. It is evident that **γ**$_{sb}$ and **K**$_L$ are the endpoints of LS(**u**$_L$).

The II$_L$ curve is defined as the union of the following *loci* (Figure 3b,c): the curve $R_1(\mathbf{\gamma}_{sp})$, the reach of the curve $S_1(\mathbf{\gamma}_{sp})$ between the states **γ**$_{sp}$ and $\mathbf{\gamma}_{sp}^\#$, the reach of the curve CD(**γ**$_{sp}$) between the states **γ**$_{sb}$ and $\mathbf{\gamma}_{sp}^\#$, the locus LS(**u**$_L$), and the reach of the curve $T_1(\mathbf{u}_L)$ with $u < 0$. The II$_L$ curve admits the particular cases II$_{L,a}$ and II$_{L,b}$, which are characterised by $u_L \leq 0$ and $u_L > 0$, respectively.

In the case II$_{L,a}$, plotted in Figure 3b for the state **u**$_L$ of Row 2 in Table 3, the state **u**$_L$ itself lies on the curve L-M below the axis $u = 0$. The points of LS(**u**$_L$) are all images of $R_1(\mathbf{u}_L)$ between **γ** and **K**$_L$.

In the case II$_{L,b}$, plotted in Figure 3c for the initial conditions of Row 3 in Table 3, the state $\mathbf{u}_L$ does not lie on the L-M curve, while the subcritical state $\mathbf{u}_L^0$, which is connected to $\mathbf{u}_L$ by SW, lies on LS($\mathbf{u}_L$). The points of LS($\mathbf{u}_L$) between $\gamma_{sb}$ and $\mathbf{u}_L^0$ are images of $R_1(\mathbf{u}_L)$ between $\gamma$ and $\mathbf{u}_L$, while the points of LS($\mathbf{u}_L$) between $\mathbf{u}_L^0$ and $\mathbf{K}_L$ are images of $S_1(\mathbf{u}_L)$ between $\mathbf{u}_L$ and $\mathbf{K}_L$.

In all the aforementioned cases, the II$_L$ curve is continuous and strictly decreasing in the ($h$, $u$) plane, and its supremum $\mathbf{I} = \begin{pmatrix} 0 & u_{\gamma_{sp}} + 2\sqrt{gh_{\gamma_{sp}}} \end{pmatrix}^T$ coincides with the dry bed state on $R_1(\gamma_{sp})$.

***III$_L$ curve***. In this case, which is characterized by $F_L > 1$, the left state $\mathbf{u}_L$ lies above the curve $C^+$ of the ($h$, $u$) plane. This implies that it is possible to consider the special subcritical state $\mathbf{u}_1 = \mathbf{u}_L^\# = \begin{pmatrix} h_L^\# & h_L^\# u_L^\# \end{pmatrix}^T$, immediately to the left of the width discontinuity, which is connected to $\mathbf{u}_L$ by means of the hydraulic jump. In addition, it is possible to define the following special $\mathbf{u}_2$ states:

- the supercritical state $\mathbf{u}_L^0 = \begin{pmatrix} h_L^0 & h_L^0 u_L^0 \end{pmatrix}^T$, connected to $\mathbf{u}_1 = \mathbf{u}_L$ by means of the standing wave SWc;

- the subcritical state $\mathbf{u}_L^{\#0} = \begin{pmatrix} h_L^{\#0} & h_L^{\#0} u_L^{\#0} \end{pmatrix}^T$, connected to $\mathbf{u}_1 = \mathbf{u}_L^\#$ by means of the standing wave SWa;

- the subcritical state $\mathbf{u}_L^{0\#} = \begin{pmatrix} h_L^{0\#} & h_L^{0\#} u_L^{0\#} \end{pmatrix}^T$, connected to $\mathbf{u}_L^0$ by means of the hydraulic jump.

From the definition, it is evident that the states $\mathbf{u}_L^{0\#}$ and $\mathbf{u}_L^{\#0}$ lie on the $CD(\mathbf{u}_L^0)$ curve, and it follows that the arc of the curve $CD(\mathbf{u}_L^0)$ between $\mathbf{u}_L^{0\#}$ and $\mathbf{u}_L^{\#0}$ represents the locus of the subcritical states $\mathbf{u}_2$ that can be connected to $\mathbf{u}_1 = \mathbf{u}_L$ by means of a SWb wave. Finally, it is possible to consider the locus LS($\mathbf{u}_L$) of the subcritical states $\mathbf{u}_2 = \mathbf{u}_1^0$ that are connected by means of the standing wave SWa to the subcritical states of $\mathbf{u}_1 \in S_1(\mathbf{u}_L)$ between $\mathbf{u}_L^\#$ and $\mathbf{K}_L$. It is evident that $\mathbf{u}_L^{\#0}$ and $\mathbf{K}_L$ are the endpoints of LS($\mathbf{u}_L$).

An example III$_L$ curve is plotted in Figure 3d for the state $\mathbf{u}_L$ of Row 4 in Table 3. The III$_L$ curve is defined as the union of the following *loci*: the curve $R_1(\mathbf{u}_L^0)$, the reach of the curve $S_1(\mathbf{u}_L^0)$ between the states $\mathbf{u}_L^0$ and $\mathbf{u}_L^{0\#}$, the reach of the curve $CD(\mathbf{u}_L^0)$ between the states $\mathbf{u}_L^{0\#}$ and $\mathbf{u}_L^{\#0}$, the locus LS($\mathbf{u}_L$), and the reach of the curve $T_1(\mathbf{u}_L)$ with $u < 0$. The III$_L$ curve is continuous and strictly decreasing, and its supremum $\mathbf{I} = \begin{pmatrix} 0 & u_L^0 + 2\sqrt{gh_L^0} \end{pmatrix}^T$ coincides with the dry bed state on $R_1(\mathbf{u}_L^0)$.

### 3.2.2 R-M curves

Four types of R-M curves, namely I$_R$, II$_R$, III$_R$, and IV$_R$, are possible, depending on the value of the Froude number $F_R = u_R / \sqrt{gh_R}$ corresponding to the state $\mathbf{u}_R$ (see Table 2, rows 4-7).

In the case $F_R > 2$ (I$_R$ curve), there is no state $\mathbf{u} \in T_2^B(\mathbf{u}_R)$ such that $u \leq 0$. This condition, which is symmetrical to that of the I$_L$ curve, implies that the state $\mathbf{u}_R$ cannot interact with the width discontinuity to its left by means of an elementary wave. Conversely, the condition $F_R \leq 2$ (II$_R$, III$_R$, and IV$_R$ curves) implies that $\mathbf{u}_R$ may be directly connected to a state $\mathbf{u}_2$ to the right of the width discontinuity by means of $T_2^B(\mathbf{u}_R)$. It is evident that the condition $F_R \leq 2$ allows considering the special state $\mathbf{K}_R = \begin{pmatrix} h_{K_R} & 0 \end{pmatrix}^T \in T_2^B(\mathbf{u}_R)$ with null velocity.

*I$_R$ curve*. For $F_R > 2$, the R-M curve coincides with $T_2^B(\mathbf{u}_R)$ and lies entirely above the axis $u = 0$ in the ($h$, $u$) plane. An example I$_R$ curve is plotted in Figure 4a for the state $\mathbf{u}_R$ of Row 1 in Table 4. It is evident that the I$_R$ curve is continuous and strictly increasing, and its infimum $\mathbf{J} = \begin{pmatrix} 0 & u_R - 2\sqrt{gh_R} \end{pmatrix}^T$ coincides with the dry bed state on $R_2^B(\mathbf{u}_R)$.

[Insert Figure 4 about here]

***II$_R$ curve***. In this R-M curve (see Figures 4b, 4c, and 4d), which is characterized by $F_R \in ]-K_{sp}(B_L/B_R), 2]$, the right state $\mathbf{u}_R$ may interact with the contraction. The special state $\boldsymbol{\alpha}_{sb} = \begin{pmatrix} h_{\alpha_{sb}} & h_{\alpha_{sb}} u_{\alpha_{sb}} \end{pmatrix}^T$ is defined as the intersection between the curve $T_2^B(\mathbf{u}_R)$ and the curve $C_{sb}^-$ of the (h, u) plane. In addition to $\boldsymbol{\alpha}_{sb}$, it is possible to consider the critical state $\mathbf{u}_1 = \boldsymbol{\alpha} = \begin{pmatrix} h_\alpha & h_\alpha u_\alpha \end{pmatrix}^T$ immediately to the left of the width discontinuity, which is connected to $\mathbf{u}_2 = \boldsymbol{\alpha}_{sb}$ by means of the standing wave SWd. By definition, $\boldsymbol{\alpha}$ lies on $C^-$. Finally, it is possible to consider the locus RS($\mathbf{u}_R$) of the subcritical states $\mathbf{u}_1 = \mathbf{u}_2^0$ that are connected by means of the standing wave SWd to the subcritical states $\mathbf{u}_2 \in T_2^B(\mathbf{u}_R)$ between $\boldsymbol{\alpha}_{sb}$ and $\mathbf{K}_R$. It is evident that $\boldsymbol{\alpha}$ and $\mathbf{K}_R$ are the endpoints of RS($\mathbf{u}_R$).

The II$_R$ curve is defined as the union of the following *loci* (see Figures 4b, 4c, and 4d): the curve $R_2^B(\boldsymbol{\alpha})$, the locus RS($\mathbf{u}_R$), and the reach of the curve $T_2^B(\mathbf{u}_R)$ with $u > 0$. The II$_R$ curve admits the particular cases II$_{R,a}$, II$_{R,b}$, and II$_{R,b}$, which differ by the position of $\mathbf{u}_R$ in the (h, u) plane.

In the case II$_{R,a}$, plotted in Figure 4b for the state $\mathbf{u}_R$ of Row 2 in Table 4, the state $\mathbf{u}_R$ itself lies on the curve R-M because $u_R > 0$, and the points of RS($\mathbf{u}_R$) are images of the states along $R_2^B(\mathbf{u}_R)$ between $\boldsymbol{\alpha}_{sb}$ and $\mathbf{K}_R$.

In the case II$_{R,b}$, plotted in Figure 4c for the state $\mathbf{u}_R$ of Row 3 in Table 4, the state subcritical state $\mathbf{u}_R$ lies between $C_{sb}^-$ and $u = 0$, and has an image $\mathbf{u}_R^0$ on the RS($\mathbf{u}_R$) curve, because $-K_{sb}(B_L/B_R) \leq F_R \leq 0$. The points of RS($\mathbf{u}_R$) between $\boldsymbol{\alpha}$ and $\mathbf{u}_R^0$ are images of $R_2^B(\mathbf{u}_R)$ between $\boldsymbol{\alpha}_{sb}$ and $\mathbf{u}_R$, while the points of RS($\mathbf{u}_R$) between $\mathbf{u}_R^0$ and $\mathbf{K}_R$ are images of $S_2^B(\mathbf{u}_R)$ between $\mathbf{u}_R$ and $\mathbf{K}_R$.

Finally, in the case II$_{R,c}$, plotted in Figure 4d for the state $\mathbf{u}_R$ of Row 4 in Table 4, the state $\mathbf{u}_R$ lies between the curves $C_{sb}^-$ and $C_{sp}^-$, and the points on $T_2^B(\mathbf{u}_R)$ in its neighbourhood have not sufficient energy to pass through the contraction because $-K_{sp}(B_L/B_R) < F_R < -K_{sb}(B_L/B_R)$. In this case, the points of RS($\mathbf{u}_R$) are images of the states along $S_2^B(\mathbf{u}_R)$ between $\alpha_{sb}$ and $\mathbf{K}_R$.

The II$_R$ curve is continuous and strictly increasing, and its infimum $\mathbf{J} = \begin{pmatrix} 0 & u_\alpha - 2\sqrt{gh_\alpha} \end{pmatrix}^T$ coincides with the dry bed state on $R_2^B(\boldsymbol{\alpha})$.

***III$_R$ curve***. An example of this locus, which is characterized by $F_R \in \,]\text{-}K_{jump}(B_L/B_R), -K_{sp}(B_L/B_R)]$, is represented in Figure 4e for the state $\mathbf{u}_R$ of Row 5 in Table 4. The special states $\boldsymbol{\alpha}_{sb}$ and $\boldsymbol{\alpha}$, and the locus RS($\mathbf{u}_R$), are defined as in the case of the II$_{R,c}$ curve. In the present case, the supercritical state $\mathbf{u}_R$ has sufficient energy to pass through the contraction because $F_R \leq -K_{sp}(B_L/B_R)$, and it is possible to consider the following special $\mathbf{u}_1$ states:

- the supercritical state $\mathbf{u}_R^0 = \begin{pmatrix} h_R^0 & h_R^0 u_R^0 \end{pmatrix}^T$, connected to $\mathbf{u}_2 = \mathbf{u}_R$ by means of the standing wave SWe;

- the subcritical state $\mathbf{u}_R^{0\#} = \begin{pmatrix} h_R^{0\#} & h_R^{0\#} u_R^{0\#} \end{pmatrix}^T$ on the curve CD($\mathbf{u}_R^0$), connected to $\mathbf{u}_R^0$ by means of the hydraulic jump;

- the critical state $\boldsymbol{\beta} = \begin{pmatrix} h_\beta & h_\beta u_\beta \end{pmatrix}^T$ at the intersection between the curve CD($\mathbf{u}_R^0$) and the curve $C^-$.

Contrary to all the preceding cases, the III$_R$ locus is not monotone, and it consists of three strictly increasing branches (III$_{R,1}$, III$_{R,2}$, and III$_{R,3}$), namely (Figure 4e):

- the branch III$_{R,1}$ is the union of $R_2^B(\boldsymbol{\alpha})$, RS($\mathbf{u}_R$), and the reach of the curve $T_2^B(\mathbf{u}_R)$ with $u > 0$, with infimum $\mathbf{J}_1 = \begin{pmatrix} 0 & u_\alpha - 2\sqrt{gh_\alpha} \end{pmatrix}^T$ coinciding with the dry bed state on $R_2^B(\boldsymbol{\alpha})$;

- the branch III$_{R,2}$ is the union of the curve CD($\mathbf{u}_R^0$) between $\boldsymbol{\beta}$ and $\mathbf{u}_R^{0\#}$, and the curve $R_2^B(\boldsymbol{\beta})$; the corresponding infimum $\mathbf{J}_2 = \begin{pmatrix} 0 & u_\beta - 2\sqrt{gh_\beta} \end{pmatrix}^T$ coincides with the dry bed state on $R_2^B(\boldsymbol{\beta})$;

- the branch III$_{R,3}$ is the union of $R_2^B(\mathbf{u}_R^0)$ and the reach of $S_2^B(\mathbf{u}_R^0)$ between $\mathbf{u}_R^0$ and $\mathbf{u}_R^{0\#}$, with infimum $\mathbf{J}_3 = \begin{pmatrix} 0 & u_R^0 - 2\sqrt{gh_R^0} \end{pmatrix}^T$ coinciding with the dry bed state on $R_2^B(\mathbf{u}_R^0)$.

Note that the state $\mathbf{u}_R$ is very close to but it does not lie on the R-M curve in Figure 4e.

***IV$_R$ curve.*** Similarly to the III$_R$ case, it is possible to define the supercritical state $\mathbf{u}_1 = \mathbf{u}_R^0$ and the subcritical state $\mathbf{u}_1 = \mathbf{u}_R^{0\#}$ because $F_R < -K_{sp}(B_L/B_R)$. Moreover, the condition $F_R \leq -K_{jump}(B_L/B_R)$ allows considering the following special states:

- the subcritical state $\mathbf{u}_2 = \mathbf{u}_R^\# = \begin{pmatrix} h_R^\# & h_R^\# u_R^\# \end{pmatrix}^T$, which is connected to $\mathbf{u}_R$ by means of the hydraulic jump;

- the subcritical state $\mathbf{u}_1 = \mathbf{u}_R^{\#0} = \begin{pmatrix} h_R^{\#0} & h_R^{\#0} u_R^{\#0} \end{pmatrix}^T$, immediately to the left of the width discontinuity, which is connected to $\mathbf{u}_2 = \mathbf{u}_R^\#$ by means of the standing wave SWd.

From the definitions, it is evident that $\mathbf{u}_R^{0\#}$ and $\mathbf{u}_R^{\#0}$ lie on the $CD(\mathbf{u}_R^0)$ curve, while the arc of the curve $CD(\mathbf{u}_R^0)$ between $\mathbf{u}_R^{0\#}$ and $\mathbf{u}_R^{\#0}$ represents the locus of the subcritical states $\mathbf{u}_1$ that can be connected to $\mathbf{u}_2 = \mathbf{u}_R$ by means of a SWf wave. Finally, the curve RS($\mathbf{u}_R$) is defined as the locus of the subcritical states $\mathbf{u}_1$ that are connected by means of the standing wave SWd to the subcritical states of $\mathbf{u}_2 \in S_2^B(\mathbf{u}_R)$ between $\mathbf{u}_R^\#$ and $\mathbf{K}_R$. The endpoints of RS($\mathbf{u}_R$) are $\mathbf{u}_R^{\#0}$ and $\mathbf{K}_R$.

Similarly to the III$_R$ locus, IV$_R$ is not a monotone curve, but it consists of three distinct branches (IV$_{R,1}$, IV$_{R,2}$, and IV$_{R,3}$) which are strictly increasing (see Figure 4f, where an example is plotted for the state $\mathbf{u}_R$ of Row 6 in Table 4):

- the branch IV$_{R,1}$ is the locus RS($\mathbf{u}_R$), and the reach of the curve $T_2^B(\mathbf{u}_R)$ with $u > 0$;

- the branch IV$_{R,2}$ is the reach of the curve CD($\mathbf{u}_R^0$) between $\mathbf{u}_R^{\#0}$ and $\mathbf{u}_R^{0\#}$;

- the branch IV$_{R,3}$ is the union of $R_2^B(\mathbf{u}_R^0)$ and the reach of $S_2^B(\mathbf{u}_R^0)$ between $\mathbf{u}_R^0$ and $\mathbf{u}_R^{0\#}$.

It is evident that the infimum $\mathbf{J} = \begin{pmatrix} 0 & u_R^0 - 2\sqrt{gh_R^0} \end{pmatrix}^T$ of the curve IV$_R$ coincides with the dry bed state on $R_2^B(\mathbf{u}_R^0)$. Notably, the state $\mathbf{u}_R$ does not lie on the IV$_R$ locus.

### 3.3 Example solutions

In the following, some Riemann exact solution examples are presented, in order to elucidate the use and the meaning of the L-M and R-M curves. All the example solutions are referred to the aspect ratio $B_L/B_R = 0.60$, but the procedure for the solution construction is totally general.

For each example, the L-M and R-M curves corresponding to the initial states $\mathbf{u}_L$ and $\mathbf{u}_R$ are constructed, and the intersection $\mathbf{u}_M$ is determined. If the velocity $u_M$ corresponding to $\mathbf{u}_M$ is positive or null, the structure of the Riemann solution is deduced from the curve L-M. Conversely, the structure of the Riemann solution is deduced from the R-M curve when $u_M < 0$. If no intersection is present, a dry bed is formed.

*Example* 1. The initial conditions for this example are contained in Table 5, row 1. The L-M curve corresponding to the state $\mathbf{u}_L$ is of the I$_L$ type (see Figure 3a), while the R-M curve corresponding to the state $\mathbf{u}_R$ is of the II$_{R,a}$ type (Figure 4b). The intersection of the two curves is the state $\mathbf{u}_M$ (Table 6, row 1), which is represented in the (*h*, *u*) plane of Figure 5a. The velocity $u_M$ is negative, implying that the flow through the geometric discontinuity is from right to left ($u_1 < 0$, $u_2 < 0$) and that the structure of the Riemann solution is deduced from the R-M curve. The state $\mathbf{u}_M$ lies on the $R_2^B(\boldsymbol{\alpha})$ reach of the curve II$_{R,a}$, and this means that (compare with Figure 4b):

- the state $\mathbf{u}_{mid}$ coincides with $\mathbf{u}_M$, and it is connected to $\mathbf{u}_1 = \boldsymbol{\alpha}$ by means of $R_2^B(\mathbf{u}_1) = R_2^B(\boldsymbol{\alpha})$;

- the state $\mathbf{u}_2$ coincides with $\boldsymbol{\alpha}_{sb}$ and is connected to $\mathbf{u}_1$ by a SWd wave;

- the state $\mathbf{u}_2$ is connected to $\mathbf{u}_R$ by means of $R_2^B(\mathbf{u}_R)$, because $\boldsymbol{\alpha}_{sb}$ lies on $R_2^B(\mathbf{u}_R)$;

- the solution structure falls in the SC4- class.

Finally, the state $\mathbf{u}_M$ lies on the $R_1(\mathbf{u}_L)$ reach of the curve $I_L$, and this implies that $\mathbf{u}_{mid}$ is connected to $\mathbf{u}_L$ by means of $R_1(\mathbf{u}_L)$. For this example, the solution structure is sketched in Table 7, row 1, while the corresponding solution (flow depth at time $t = 5s$) is plotted in Figure 5b. In Figure 5b, a black arrow depicts the direction of the flow through the geometric discontinuity.

[Insert Figure 5 about here]

*Example* 2. The corresponding initial conditions are contained in row 2 of Table 5. The L-M curve is a $II_{L,a}$ curve (Figure 3b), while the R-M curve is a $II_{R,b}$ curve (Figure 4c). The intersection $\mathbf{u}_M$ of Figure 5c is characterized by negative velocity (Table 6, row 2) and lies on the $RS(\mathbf{u}_R)$ reach of the $II_{R,b}$ curve, implying that (compare with Figure 4c):

- the state $\mathbf{u}_1$ coincides with $\mathbf{u}_M$;

- the state $\mathbf{u}_2$ is connected to $\mathbf{u}_1$ by a SWd wave, and it is easily computed using the conditions of energy and discharge invariance;

- the state $\mathbf{u}_2$ is connected to $\mathbf{u}_R$ by means of $R_2^B(\mathbf{u}_R)$, because $\mathbf{u}_M$ lies to the left of $\mathbf{u}_R^0$ (see the description of the curve $II_{R,b}$ in section 3.2.2);

- the solution structure falls in the SC3- class.

The state $\mathbf{u}_M$ lies on the $R_1(\mathbf{u}_L)$ reach of the curve $II_{L,a}$, implying that $\mathbf{u}_I$ is connected to $\mathbf{u}_L$ by means of a rarefaction contained into the first characteristic field. The corresponding solution structure is sketched in Table 7, row 2, while the solution (flow depth at time $t = 5$s) is plotted in Figure 5d.

*Example* 3. The initial conditions for this example are contained in row 3 of Table 5. The L-M curve is a II$_{L,b}$ curve (Figure 3c), while the R-M curve is a II$_{R,a}$ curve (Figure 4b). The intersection $\mathbf{u}_M$ of Figure 5e is characterized by positive velocity (Table 6, row 3), implying that the flow through the geometric discontinuity is from left to right ($u_1 > 0$, $u_2 > 0$) and that the structure of the Riemann solution can be deduced from the L-M curve. In particular, the state $\mathbf{u}_M$ lies on the $S_1(\gamma_{sp})$ reach of the curve II$_{L,b}$, and this means that (compare with Figure 3c):

- the state $\mathbf{u}_{mid}$ coincides with $\mathbf{u}_M$, and it is connected to $\mathbf{u}_1 = \gamma$ by means of $S_1(\mathbf{u}_1) = S_1(\gamma_{sp})$;
- the state $\mathbf{u}_2$ coincides with $\gamma_{sp}$ and is connected to $\mathbf{u}_1$ by a SWc wave;
- the state $\mathbf{u}_1$ is connected to $\mathbf{u}_L$ by means of $R_1(\mathbf{u}_L)$, because $\gamma$ lies on $R_1(\mathbf{u}_L)$;
- the solution structure falls in the SC4+ class.

The state $\mathbf{u}_M$ lies on the $R_2^B(\mathbf{u}_R)$ reach of the curve II$_{R,a}$, implying that $\mathbf{u}_{mid}$ is connected to $\mathbf{u}_R$ by means of $R_2^B(\mathbf{u}_R)$. The corresponding solution structure is sketched in Table 7, row 3, while the solution (flow depth at time $t = 5$s) is plotted in Figure 5f.

*Example* 4. For this example, whose initial conditions are contained in row 4 of Table 5, the L-M curve is a II$_{L,b}$ curve (Figure 3c), while the R-M curve is a II$_{R,b}$ curve (Figure 4c). The intersection $\mathbf{u}_M$ of Figure 6a, which is characterized by positive velocity (Table 6, row 4), lies on the LS($\mathbf{u}_L$) reach of the curve II$_{L,b}$ to the right of $\mathbf{u}_L^0$. This implies that (compare with Figure 3c):

- the state $\mathbf{u}_2$ coincides with $\mathbf{u}_M$;

- the state $\mathbf{u}_1$ is connected to the state $\mathbf{u}_2$ by means of a SWa wave, and it is easily computed using the conditions of energy and discharge invariance;
- the state $\mathbf{u}_1$ is connected to $\mathbf{u}_L$ by means of $S_1(\mathbf{u}_L)$, because $\mathbf{u}_M$ lies on LS($\mathbf{u}_L$) to the right of $\mathbf{u}_L^0$ (see the description of the II$_{L,b}$ curve in section 3.2.1);
- the solution structure falls in the SC3+ class.

Finally, the state $\mathbf{u}_M$ lies on the $S_2^B(\mathbf{u}_R)$ reach of the curve II$_{R,b}$, implying that $\mathbf{u}_2$ is connected to $\mathbf{u}_R$ by means of $S_2^B(\mathbf{u}_R)$. The corresponding solution structure is sketched in Table 7, row 4, while the solution (flow depth at time $t = 5$s) is plotted in Figure 6b.

[Insert Figure 6 about here]

*Example* 5. The corresponding initial conditions are contained in Table 5, row 5. The L-M curve is a II$_{L,b}$ curve (Figure 3c), while the R-M curve is a II$_{R,c}$ curve (Figure 4d). The intersection $\mathbf{u}_M$ of Figure 6c, which is characterized by negative velocity (Table 6, row 5), lies on the RS($\mathbf{u}_R$) reach of the curve II$_{R,c}$, implying that:
- the state $\mathbf{u}_1$ coincides with $\mathbf{u}_M$;
- the state $\mathbf{u}_2$ is connected to the state $\mathbf{u}_1$ by means of a SWd wave, and it is easily computed using the conditions of energy and discharge invariance;
- the state $\mathbf{u}_2$ is connected to $\mathbf{u}_R$ by means of $S_2^B(\mathbf{u}_R)$, because all the states on RS($\mathbf{u}_R$) are images of points along $S_2^B(\mathbf{u}_R)$.
- the solution structure falls in the SC3- class.

The state $\mathbf{u}_M$ lies on the $S_1(\mathbf{u}_L)$ reach of the curve II$_{L,b}$, implying that $\mathbf{u}_1$ is connected to $\mathbf{u}_L$ by means of $S_1(\mathbf{u}_L)$. The solution structure is sketched in Table 7, row 5, while the solution (flow depth at time $t = 5$s) is plotted in Figure 6d.

*Example* 6. The corresponding initial conditions are contained in Table 5, row 6. The L-M curve is a $III_L$ curve (Figure 3d), while the R-M curve is a $II_{R,b}$ curve (Figure 4c). The intersection $\mathbf{u}_M$ of Figure 6e has positive velocity (Table 6, row 6) and lies on the $CD(\mathbf{u}_L^0)$ reach of the curve $III_L$, implying that:

- the state $\mathbf{u}_2$ coincides with $\mathbf{u}_M$;
- the state $\mathbf{u}_1$ coincides with $\mathbf{u}_L$, and it is connected to $\mathbf{u}_2$ by means of a SWb wave;
- the solution structure falls in the $SC2_0+$ class.

Finally, the state $\mathbf{u}_M$ lies on the $S_2^B(\mathbf{u}_R)$ reach of the curve $II_{R,b}$, implying that $\mathbf{u}_2$ is connected to $\mathbf{u}_R$ by means of a shock contained into the second characteristic field. The solution structure is sketched in Table 7, row 6, while the solution (flow depth at time $t = 5$s) is plotted in Figure 6f.

*Example* 7. This case corresponds to the initial conditions of Table 5, row 7. The L-M curve is a $III_L$ curve (Figure 3d), while the R-M curve is a $II_{R,a}$ curve (Figure 4b). The intersection $\mathbf{u}_M$ of Figure 7a (see Table 6, row 7) is characterized by positive velocity and lies on the $S_1(\mathbf{u}_L^0)$ reach of the $III_L$ curve, implying that:

- the state $\mathbf{u}_1$ coincides with $\mathbf{u}_L$;
- the state $\mathbf{u}_2$ coincides with $\mathbf{u}_L^0$, and it is connected to $\mathbf{u}_1$ by means of a SWc wave;
- the state $\mathbf{u}_{mid}$ coincides with $\mathbf{u}_M$, and it is connected to $\mathbf{u}_2$ by means of $S_1(\mathbf{u}_L^0)$;
- the solution structure falls in the $SC3_0+$ class.

In addition, the state $\mathbf{u}_M$ lies on the $S_2^B(\mathbf{u}_R)$ reach of the curve $II_{R,a}$, implying that $\mathbf{u}_{mid}$ is connected to $\mathbf{u}_R$ by means of a shock contained into the second characteristic field. A sketch of the solution structure is reported in Table 7, row 7. The solution (flow depth at time $t = 5$s) is plotted in Figure 7b.

[Insert Figure 7 about here]

*Example* 8. The initial conditions for this case are contained in Table 5, row 8. The L-M curve is a $II_{L,b}$ curve (Figure 3c), while the R-M curve is a $II_{R,a}$ curve (compare with Figure 4b). The intersection $\mathbf{u}_M$ of Figure 7c (see Table 6, row 8) has positive velocity and lies on the $CD(\gamma_{sp})$ reach of the $II_{L,b}$ curve, implying that:

- the state $\mathbf{u}_1$ coincides with the critical state $\gamma$, and it is connected to $\mathbf{u}_L$ by means of the $R_1(\mathbf{u}_L)$ wave;
- the state $\mathbf{u}_2$ coincides with $\mathbf{u}_M$, and it is connected to $\mathbf{u}_1$ by means of a SWb wave;
- the solution structure falls in the SC3+ class.

In addition, the state $\mathbf{u}_M$ lies on the $R_2^B(\mathbf{u}_R)$ reach of the curve $II_{R,b}$, implying that $\mathbf{u}_2$ is connected to $\mathbf{u}_R$ by means of a rarefaction contained into the second characteristic field. A sketch of the solution structure is reported in Table 7, row 8. The solution (flow depth at time $t = 5$s) is plotted in Figure 7d.

*Example* 9. The initial conditions for this example are contained in Table 5, row 9. The L-M curve corresponding to the state $\mathbf{u}_L$ is of the $I_L$ type (compare with Figure 3a), while the R-M curve corresponding to the state $\mathbf{u}_R$ is of the $II_{R,a}$ type (Figure 4b). The two curves have no intersection (see Figure 7e), meaning that a dry bed is formed somewhere between the states $\mathbf{u}_L$ and $\mathbf{u}_R$. The dry bed state **I** on the curve L-M and the dry bed state **J** on the curve R-M are characterized by negative velocity, implying that the flow through the geometric discontinuity is from right to left ($u_1 < 0$, $u_2$

< 0) and that the structure of the Riemann solution is deduced from the R-M curve. In particular (compare with Figure 4b):

- the state $\mathbf{u}_1$ coincides with $\boldsymbol{\alpha}$;
- the state $\mathbf{u}_2$ coincides with $\boldsymbol{\alpha}_{sb}$ and it is connected to $\mathbf{u}_1$ by a SWd wave;
- the state $\mathbf{u}_2$ is connected to $\mathbf{u}_R$ by means of $R_2^B(\mathbf{u}_R)$;
- the dry bed state **J** is connected to $\mathbf{u}_1$ by means of $R_2^B(\mathbf{u}_1) = R_2^B(\boldsymbol{\alpha})$;
- the dry bed state **I** is connected to $\mathbf{u}_L$ by means of $R_1(\mathbf{u}_L)$.

It is evident that the solution structure is a particular case of the SC4- class (compare with the Example 1 Riemann problem). The solution structure is sketched in Table 7, row 9, while the corresponding solution (flow depth at time $t = 5$s) is plotted in Figure 7f.

*Example* 10. For this example, the initial conditions are contained in Table 5, row 10. The L-M curve corresponding to the state $\mathbf{u}_L$ is of the $\mathrm{II}_{L,a}$ type (Figure 3b), while the R-M curve corresponding to the state $\mathbf{u}_R$ is of the $\mathrm{III}_R$ type (Figure 4e). These two *loci* are plotted in Figure 8a, and again in Figures 8c and 8e, showing that three distinct intersections ($\mathbf{u}_{M,1}$ in Figure 8a, $\mathbf{u}_{M,2}$ in Figure 8c, $\mathbf{u}_{M,3}$ in Figure 8e) are present, corresponding to three distinct solutions of the Riemann problem.

The intersection $\mathbf{u}_{M,1}$ of Figure 8a (see the corresponding values of flow depth and velocity in Table 6, row 10) has negative velocity and lies on the $R_2^B(\boldsymbol{\alpha})$ reach of the $\mathrm{III}_R$ curve (compare with Figure 4e), implying that:

- the state $\mathbf{u}_1$ coincides with the critical state $\boldsymbol{\alpha}$;
- the state $\mathbf{u}_2$ coincides with $\boldsymbol{\alpha}_{sb}$ and it is connected to $\mathbf{u}_1$ by a SWd wave;
- the state $\mathbf{u}_{mid}$ coincides with $\mathbf{u}_{M,1}$ and it is connected to $\mathbf{u}_1$ by means of $R_2^B(\mathbf{u}_1) = R_2^B(\boldsymbol{\alpha})$;

- the state $\mathbf{u}_2$ is connected to $\mathbf{u}_R$ by means of $S_2^B(\mathbf{u}_R)$, because the state $\alpha_{sb}$ of the III$_R$ *locus* is connected to the state $\mathbf{u}_R$ by a shock;
- the solution structure falls in the SC4- class.

The state $\mathbf{u}_{M,1}$ lies on the $S_1(\mathbf{u}_L)$ reach of the curve I$_L$, and this implies that $\mathbf{u}_{mid}$ is connected to $\mathbf{u}_L$ by means of a shock. The corresponding solution structure is sketched in Table 7, row 10, while the flow depth at time $t = 5$s is plotted in Figure 8b.

The intersection $\mathbf{u}_{M,2}$ of Figure 8c (Table 6, row 11) lies on the $R_2^B(\boldsymbol{\beta})$ reach of the III$_R$ curve (compare with Figure 4e). This implies that:

- the state $\mathbf{u}_1$ coincides with the critical state $\boldsymbol{\beta}$;
- the state $\mathbf{u}_2$, which coincides with $\mathbf{u}_R$ is connected to $\mathbf{u}_1$ by a SWf wave;
- the state $\mathbf{u}_{mid}$ coincides with $\mathbf{u}_{M,2}$ and it is connected to $\mathbf{u}_1$ by means of $R_2^B(\mathbf{u}_1) = R_2^B(\boldsymbol{\beta})$;
- the solution structure falls in the SC3$_0$- class.

Finally, the state $\mathbf{u}_{M,2}$ lies on the $S_1(\mathbf{u}_L)$ reach of the curve I$_L$, and this implies that $\mathbf{u}_{mid}$ is connected to $\mathbf{u}_L$ by means of a shock. The corresponding solution structure is sketched in Table 7, row 11, while the flow depth at time $t = 5$s is plotted in Figure 8d.

The intersection $\mathbf{u}_{M,3}$ of Figure 8e (Table 6, row 12) lies on the $S_2^B(\mathbf{u}_R^0)$ reach of the III$_R$ curve (compare with Figure 4e). This implies that:

- the state $\mathbf{u}_1$ coincides with $\mathbf{u}_R^0$;
- the state $\mathbf{u}_2$, which coincides with $\mathbf{u}_R$, is connected to $\mathbf{u}_1$ by a SWe wave;
- the state $\mathbf{u}_{mid}$ coincides with $\mathbf{u}_{M,3}$ and it is connected to $\mathbf{u}_1$ by means of $S_2^B(\mathbf{u}_1) = S_2^B(\mathbf{u}_R^0)$;
- the solution structure falls in the SC3$_0$- class.

Again, the state $\mathbf{u}_{M,1}$ lies on the $S_1(\mathbf{u}_L)$ reach of the curve $I_L$, and this implies that $\mathbf{u}_{mid}$ is connected to $\mathbf{u}_L$ by means of a shock. The corresponding solution structure is sketched in Table 7, row 12, while the flow depth at time $t = 5$s is plotted in Figure 8f.

*Example* 11. The initial conditions for this example are contained in Table 5, row 11. The L-M curve corresponding to the state $\mathbf{u}_L$ is of the $III_L$ type (compare with Figure 3d), while the R-M curve corresponding to the state $\mathbf{u}_R$ is of the $IV_R$ type (see Figure 4f). The corresponding *loci* are plotted in panels a, c, and e of Figure 9. Similarly to the case of Example 10, in the Example 11 the *loci* L-M and R-M have three distinct intersections ($\mathbf{u}_{M,1}$ in Figure 9a, $\mathbf{u}_{M,2}$ in Figure 9c, $\mathbf{u}_{M,3}$ in Figure 9e), which correspond to three distinct solutions of the Riemann problem.

The intersection $\mathbf{u}_{M,1}$ of Figure 9a (values of flow depth and velocity in Table 6, row 13) has negative velocity and lies on the $RS(\mathbf{u}_R)$ reach of the $IV_R$ curve (compare with Figure 4f), implying that:

- the state $\mathbf{u}_1$ coincides with the state $\mathbf{u}_{M,1}$, and $\mathbf{u}_2$ is connected to $\mathbf{u}_1$ by a SWd wave;
- the state $\mathbf{u}_2$ is connected to $\mathbf{u}_R$ by a shock contained into the second characteristic field, because the points of $RS(\mathbf{u}_R)$ are images of states along $S_2^B(\mathbf{u}_R)$;
- the solution structure falls in the SC3- class.

Note that the state $\mathbf{u}_{M,1}$ lies on the $S_1(\mathbf{u}_L)$ reach of the curve $III_L$, implying that $\mathbf{u}_1$ is connected to $\mathbf{u}_L$ by means of a shock contained into the first characteristic field. The sketch of the solution structure is reported in Table 7, row 13, while the flow depth at time $t = 5$s is plotted in Figure 9b.

The intersection $\mathbf{u}_{M,2}$ of Figure 9c (values in Table 6, row 14) lies on the $CD(\mathbf{u}_R^0)$ reach of the $IV_R$ curve (compare with Figure 4f), implying that:

- the state $\mathbf{u}_1$ coincides with the state $\mathbf{u}_{M,2}$;
- the state $\mathbf{u}_2$, which coincides with $\mathbf{u}_R$, is connected to $\mathbf{u}_1$ by a SWf wave;

- the solution structure falls in the SC2$_0$- class.

Finally, the state $\mathbf{u}_{M,2}$ lies on the $S_1(\mathbf{u}_L)$ reach of the curve III$_L$, and this implies that $\mathbf{u}_1$ is connected to $\mathbf{u}_L$ by means of a shock contained into the first characteristic field. The corresponding solution structure is sketched in Table 7, row 14, while the flow depth at time $t = 5$s is plotted in Figure 8d.

The intersection $\mathbf{u}_{M,3}$ of Figure 9e (values in Table 6, row 15) lies on the $S_2^B(\mathbf{u}_R^0)$ reach of the IV$_R$ curve (compare with Figure 4e). This implies that:

- the state $\mathbf{u}_1$ coincides with $\mathbf{u}_R^0$;
- the state $\mathbf{u}_2$, which coincides with $\mathbf{u}_R$, is connected to $\mathbf{u}_1$ by a SWe wave;
- the state $\mathbf{u}_{mid}$ coincides with $\mathbf{u}_{M,3}$ and it is connected to $\mathbf{u}_1$ by means of
$$S_2^B(\mathbf{u}_1) = S_2^B(\mathbf{u}_R^0);$$
- the solution structure falls in the SC3$_0$- class.

The state $\mathbf{u}_{M,1}$ lies on the $S_1(\mathbf{u}_L)$ reach of the curve I$_L$, and this implies that $\mathbf{u}_{mid}$ is connected to $\mathbf{u}_L$ by means of a shock contained into the first characteristic field. The corresponding solution structure is sketched in Table 7, row 15, while the flow depth at time $t = 5$s is plotted in Figure 9f.

### 3.4 Global characteristics of the solutions

For practical uses, it is important to know if the solution of the Riemann problem at width jumps always exists and if it is unique or not. In the present sub-section, it will be demonstrated that the solution to the Shallow water Equations Riemann problem at width jumps always exists. In addition, it will be demonstrated that, depending on the left and right states $\mathbf{u}_L$ and $\mathbf{u}_R$, the solution may be unique or triple. The corresponding reasoning is based on the monotonicity properties of L-M and R-M curves.

Before proving the global characteristics of the solutions, we observe that the L-M *loci* ($I_L$, $II_L$, and $III_L$) can be represented as continuous strictly decreasing functions in the form $u = f_{LM}(h)$ (see Figure 3) in the interval $[0, +\infty[$. For $h = 0$, these functions attain the value $u_I = f_{LM}(0)$, which is the velocity corresponding to the dry bed state **I**, while they tend to $-\infty$ for $h \to +\infty$. First we prove the following

**Lemma 1**. *If $F_R > -K_{sp}(B_L/B_R)$, the solution of the Shallow water Equations Riemann problem at jump widths with $B_L < B_R$ always exists and it is unique.*

*Proof.* The R-M *loci* $I_R$ and $II_R$, which are characterized by $F_R > -K_{sp}(B_L/B_R)$, can be represented in the plane $(h, u)$ by continuous strictly increasing functions with form $u = f_{RM}(h)$ in the interval $[0, +\infty[$ (see Figure 4a,b,c,d). For $h = 0$, these functions attain the value $f_{RM}(0) = u_J$, which is the velocity corresponding to the dry bed state **J**, while they tend to $+\infty$ for $h \to +\infty$.

Consider the difference $g(h) = f_{LM}(h) - f_{RM}(h)$ in the interval $[0, +\infty[$. The function $u = g(h)$ is continuous and strictly decreasing, attaining the value $g(0) = u_I - u_J$ in $h = 0$, and tending to $-\infty$ for $h \to +\infty$. Two cases are now possible:

a) If $u_I - u_J \geq 0$, the Intermediate Value Theorem applies, demonstrating that the solution of the equation $g(h) = 0$ exists and it is unique. This implies that the intersection of L-M and R-M exists and it is unique, guaranteeing that the solution of the Riemann problem exists and it is unique.

b) If $u_I - u_J < 0$, the *loci* L-M and R-M do not intersect because the solution of the equation $g(h) = 0$ does not exist. Also in this case, the solution of the Riemann problem exists and it is unique because it corresponds to the formation of a dry bed.

Now we prove the following

**Lemma 2**. *If $F_R \in$ ]-$K_{jump}(B_L/B_R)$, -$K_{sp}(B_L/B_R)$], the solution of the Shallow water Equations Riemann problem at jump widths with $B_L < B_R$ always exists. The solution may be unique or triple depending on $\mathbf{u}_L$.*

*Proof.* The III$_R$ locus, which is valid for $F_R \in$ ]-$K_{jump}(B_L/B_R)$, -$K_{sp}(B_L/B_R)$], consists of three branches (III$_{R,1}$, III$_{R,2}$, and III$_{R,3}$), which can be represented in the plane $(h, u)$ by:

- III$_{R,1}$: a strictly increasing function with form $u = f_{RM,1}(h)$ in the interval [0, + ∞ [, such that $f_{RM,1}(0) = u_{J,1}$ and that $f_{RM,1}(h) \to$ + ∞ for $h \to$ + ∞.

- III$_{R,2}$: a strictly increasing function with form $u = f_{RM,2}(h)$ in the interval [0, $h_R^{0\#}$ ], such that $f_{RM,2}(0) = u_{J,2}$ and that $f_{RM,2}(h_R^{0\#}) = u_R^{0\#}$.

- III$_{R,3}$: a strictly increasing function with form $u = f_{RM,3}(h)$ in the interval [0, $h_R^{0\#}$ ], such that $f_{RM,3}(0) = u_{J,3}$ and that $f_{RM,3}(h_R^{0\#}) = u_R^{0\#}$.

These functions are such that $f_{RM,1}(h) > f_{RM,2}(h) > f_{RM,3}(h)$ in [0, $h_R^{0\#}$ [, while $f_{RM,1}(h_R^{0\#}) > u_R^{0\#}$. Given the state $\mathbf{u}_L$, which determines the sign of the difference $\Delta u = f_{LM}\left(h_R^{0\#}\right) - u_R^{0\#}$, two distinct cases are possible:

a) If $\Delta u > 0$, consider the difference function $g_1(h) = f_{LM}(h) - f_{RM,1}(h)$ in the interval [0, + ∞[. The same reasoning of *Lemma* 1 proves that the solution of the Riemann problem exists and it is unique.

b) If $\Delta u \leq 0$, consider the difference functions $g_1(h) = f_{LM}(h) - f_{RM,1}(h)$, $g_1(h) = f_{LM}(h) - f_{RM,2}(h)$, and $g_1(h) = f_{LM}(h) - f_{RM,3}(h)$, in the interval [0, $h_R^{0\#}$ ]. The iteration of the reasoning from *Lemma* 1 proves that the solution exists and it is triple. Of course, two of these solutions coincide in the case $\Delta u = 0$.

**Lemma 3**. *If $F_R \leq -K_{jump}(B_L/B_R)$, the solution of the Shallow water Equations Riemann problem at jump widths with $B_L < B_R$ always exists. The solution may be unique or triple depending on $\mathbf{u}_L$.*

*Proof.* The IV$_R$ locus, which is valid for $F_R \leq -K_{jump}(B_L/B_R)$, consists of three branches (IV$_{R,1}$, IV$_{R,2}$, and IV$_{R,3}$), which can be represented in the plane $(h, u)$ by:

- IV$_{R,1}$: a strictly increasing function with form $u = f_{RM,1}(h)$ in the interval $[h_R^{\#0}, +\infty[$, such that $f_{RM,1}(h_R^{\#0}) = u_R^{\#0}$ and that $f_{RM,1}(h) \to +\infty$ for $h \to +\infty$.

- IV$_{R,2}$: a strictly increasing function with form $u = f_{RM,2}(h)$ in the interval $[h_R^{\#0}, h_R^{0\#}]$, such that $f_{RM,2}(h_R^{\#0}) = u_R^{\#0}$ and that $f_{RM,2}(h_R^{0\#}) = u_R^{0\#}$.

- IV$_{R,3}$: a strictly increasing function with form $u = f_{RM,3}(h)$ in the interval $[0, h_R^{0\#}]$, such that $f_{RM,3}(0) = u_J$ and that $f_{RM,3}(h_R^{0\#}) = u_R^{0\#}$.

These functions are such that $f_{RM,1}(h) > f_{RM,2}(h) > f_{RM,3}(h)$ in $]h_R^{\#0}, h_R^{0\#}[$, $f_{RM,1}(h_R^{0\#}) > u_R^{0\#}$, $f_{RM,3}(h_R^{\#0}) < u_R^{\#0}$, with $u_R^{0\#} > u_R^{\#0}$. Given the state $\mathbf{u}_L$, which determines the sign of the differences $\Delta u_1 = f_{LM}(h_R^{0\#}) - u_R^{0\#}$ and $\Delta u_2 = f_{LM}(h_R^{\#0}) - u_R^{\#0}$, three distinct cases are possible:

a) If $\Delta u_1 > 0$, consider the difference function $g_1(h) = f_{LM}(h) - f_{RM,1}(h)$ in the interval $[h_R^{\#0}, +\infty[$. The same reasoning of *Lemma* 1 proves that the solution of the Riemann problem exists and it is unique.

b) If $\Delta u_2 < 0$, consider the difference function $g_3(h) = f_{LM}(h) - f_{RM,3}(h)$ in the interval $[0, h_R^{\#0}]$. The same reasoning of *Lemma* 1 proves that the solution of the Riemann problem exists and it is unique.

c) In the other cases, consider the difference functions $g_1(h) = f_{LM}(h) - f_{RM,1}(h)$, $g_2(h) = f_{LM}(h) - f_{RM,2}(h)$, and $g_3(h) = f_{LM}(h) - f_{RM,3}(h)$, in the interval $[h_R^{\#0}, h_R^{0\#}]$. The iteration of the reasoning from *Lemma* 1 proves that the solution exists and it is triple. Of course, two of these solutions coincide in the case $\Delta u_1 = 0$ or $\Delta u_2 = 0$.

In conclusion, it can be stated the following

**Theorem 1**. *The solution of the Shallow water Equations Riemann problem at jump widths with $B_L < B_R$ always exists. If $F_R > -K_{sp}(B_L/B_R)$, the solution is unique. If $F_R \leq -K_{sp}(B_L/B_R)$, the solution may be unique or triple depending on $\mathbf{u}_L$.*

*Proof*. The theorem follows trivially from *Lemmas* 1 to 3

## 4. Numerical experiments

The exact solution of the Riemann problem at width discontinuities can be used as a benchmark for the evaluation of current one-dimensional Shallow water Equations numerical models, and an example of such a comparison is presented here. Due to the formal identity between one-dimensional variable-width Shallow water Equations and one-dimensional Porous Shallow water Equations (Guinot and Soares-Frazão 2006, Sanders et al. 2008), the numerical model by Cozzolino et al. (2018b) is considered, after the substitution of the porosity symbol with the channel width symbol.

### 4.1 Numerical model description

The numerical model approximates the solution of the system of Eq. (1) by means of the following first-order Finite Volume scheme

(7) $B_i\mathbf{u}_i^{n+1} - B_i\mathbf{u}_i^n = -\dfrac{\Delta t}{\Delta x}\left[B_{i+1/2}\mathbf{g}\left(\mathbf{u}_{i+1/2}^-,\mathbf{u}_{i+1/2}^+\right) - B_{i-1/2}\mathbf{g}\left(\mathbf{u}_{i-1/2}^-,\mathbf{u}_{i-1/2}^+\right)\right] + \dfrac{\Delta t}{\Delta x}\left[\mathbf{s}_{i+1/2}^- - \mathbf{s}_{i-1/2}^+\right].$

In Eq. (7), $\Delta x = x_{i+1/2} - x_{i-1/2}$ is the length of the cell $C_i = [x_{i-1/2}, x_{i+1/2}]$ with centre $x_i = 0.5\left(x_{i-1/2} + x_{i+1/2}\right)$, while

(8) $B_i = \dfrac{1}{\Delta x_i}\int_{C_i} B(x)\,dx, \quad \mathbf{u}_i^n = \left(h_i^n \quad h_i^n u_i^n\right)^T = \dfrac{1}{B_i \Delta x_i}\int_{C_i} B(x)\mathbf{u}(x,t_n)\,dx$

are the cell-averages of the width $B(x)$ and of the conserved vector $\mathbf{u}(x,t)$ at time level $t_n$. In addition, $B_{i+1/2}$ is the channel width at the cell interfaces; $\mathbf{g}(\mathbf{v}, \mathbf{w})$ is a numerical flux vector corresponding to the unit-width one-dimensional Shallow water Equations; $\mathbf{u}_{i+1/2}^- = \left(h_{i+1/2}^- \quad h_{i+1/2}^- u_{i+1/2}^-\right)^T$ and $\mathbf{u}_{i+1/2}^+ = \left(h_{i+1/2}^+ \quad h_{i+1/2}^+ u_{i+1/2}^+\right)^T$ are the conserved variables reconstructed to the left and to the right of the $x_{i+1/2}$ interface, respectively; finally, $\mathbf{s}_{i+1/2}^-$ and $\mathbf{s}_{i-1/2}^+$ are the contributions to the cell $C_i$ of the non-conservative product $\mathbf{h}(\mathbf{u})\partial B/\partial x$ through interfaces $x_{i+1/2}$ and $x_{i-1/2}$, respectively.

Following Cozzolino et al. (2018b), the variables $\mathbf{u}_{i+1/2}^-$ and $\mathbf{u}_{i+1/2}^+$ are reconstructed from $\mathbf{u}_i^n$ and $\mathbf{u}_{i+1}^n$ in order to ensure conservation of discharge and energy through the monotonic discontinuity. Given the position $B_{i+1/2} = \min(B_i, B_{i+1})$, the reconstructed variables $\mathbf{u}_{i+1/2}^- = \left(h_{i+1/2}^- \quad B_i h_i u_i / B_{i+1/2}\right)^T$ and $\mathbf{u}_{i+1/2}^+ = \left(h_{i+1/2}^+ \quad B_{i+1} h_{i+1} u_{i+1} / B_{i+1/2}\right)^T$ satisfy the conditions

(9) $\begin{aligned} h_{i+1/2}^- + \left(B_i h_i u_i\right)^2 / \left[2g\left(B_{i+1/2} h_{i+1/2}^-\right)^2\right] &= h_i + u_i^2/(2g) \\ h_{i+1/2}^+ + \left(B_{i+1} h_{i+1} u_{i+1}\right)^2 / \left[2g\left(B_{i+1/2} h_{i+1/2}^+\right)^2\right] &= h_{i+1} + u_{i+1}^2/(2g) \end{aligned}.$

Note that this procedure is equivalent to say that the standing wave SWd is chosen when the flow to be reconstructed is subcritical, while the standing wave SWe is chosen when the flow is supercritical. A special procedure is adopted in the case that the condition of specific energy invariance expressed by Eq. (9) cannot be satisfied. Similarly, an additional special approach is used to detect the presence of a hydraulic jump through the geometric discontinuity when a rarefaction is attached to the width discontinuity (for more details, see Cozzolino et al. 2018b).

This numerical model has been extensively tested using the dam-break exact solutions contained in Cozzolino et al. (2018b), but the availability of the Riemann problem solutions allows a more complete evaluation.

## 4.2 Numerical tests

In the following numerical examples, a channel of length $L = 200$ m exhibits a width discontinuity at its centre $x = 0$ m. The channel width is $B_L = 0.60$ m for $x < 0$ while it is $B_R = 1.00$ m for $x > 0$. The Finite Volume scheme is applied to the Riemann problems of Table 5 with $\Delta x = 0.20$ m and $\Delta t = 0.005$ s, and the numerical solutions (flow depth $h$) at time $t = 5$ s are compared in Figures 10 and 11 with the corresponding exact solutions.

The numerical solution for the Riemann problem of row 1 in Table 5 is represented in Figure 10a (dashed line). This case is characterized by flow from right to left through a contraction, with the formation of a critical state immediately to the left of the discontinuity. The comparison with the exact solution (continuous line) shows that the standing wave discontinuity is captured almost exactly, while slightly less satisfactory results are obtained for the three rarefaction waves, due to the numerical viscosity of the first-order scheme. Of course, numerical viscosity can be reduced by refining the numerical grid, or by resorting to a second-order accurate scheme.

[Insert Figure 10 about here]

Similar results are obtained in Figure 10b for the Riemann problem of Table 5, row 2. This case is characterized by subcritical flow through the discontinuity, with the formation of two rarefactions on the left and on the right of the standing wave respectively. The comparison between the numerical and the exact solution (dashed and continuous line, respectively) shows that the standing wave is nicely captured, while the numerical viscosity has an influence on the moving rarefactions.

The Riemann problem of Table 5, row 3, is very interesting because it exhibits a resonant condition where a rarefaction of the first characteristic field is attached to the left of the width discontinuity. From the literature, it is well known that accurate simulation of this condition may be problematic for numerical schemes (Thanh 2013). The inspection of Figure 10c demonstrates that the numerical model (dashed line) is able to approximate the exact solution (continuous line) also in this special case.

In Figures 10d and 10e (initial conditions in Table 5, rows 4 and 5, respectively), the numerical model (dashed line) accurately captures the standing wave and the two shocks emerging from the initial discontinuity exhibited by the exact solution (continuous line).

The problem of Table 5, row 6, is very interesting because the left supercritical flow is converted into a subcritical flow through a hydraulic jump located into the geometric discontinuity, which is a condition never considered in the literature. Apparently, the numerical model is able to capture the exact solution, as shown in Figure 10f, even if the algorithm is not expressly designed for this special condition. The problem of Table 5, row 7, differs from the preceding case because the left supercritical flow is able to push the shock out of the standing wave, as shown in Figure 11a. In this case, minor differences between numerical and exact solution are due to numerical diffusion.

[Insert Figure 11 about here]

The case of Figure 11b, corresponding to the problem of Table 5, row 8, exhibits a hydraulic jump through the width discontinuity, where the supercritical flow is reverted to subcritical. The peculiarity of this case is that this phenomenon appears jointly with a rarefaction contained into the first characteristic field, which is attached to the width discontinuity. In this special condition, which was tested in Cozzolino et al. (2018b) for the dam-break case, the numerical model apparently behaves nicely. In Figure 11c, the numerical solution for the problem of Table 5, row 9, is represented. The comparison with the exact solution shows that the numerical model is able to capture the cavitation phenomenon, that is the formation of dry bed which is initially wet everywhere.

The cases of Figures 11d and 11e are special because they correspond to Riemann problems (Table 5, rows 10 and 11, respectively) exhibiting multiple solutions. For these cases, the inspection of Figures 11d and 11e shows that in both problems the numerical model captures the solution characterized by supercritical flow freely passing through the width discontinuity. The comparison with Figures 8f and 9f shows negligible differences between the exact and the numerical solution.

From these observations, it seems that the numerical model is able to capture the Riemann problem exact solution in all the cases. This conclusion is partial and it can be moderated if the plot of the discharge $Q = Bhu$ is examined. For the sake of brevity, only the Riemann problems of Table 5, rows 1, 2, 6, and 8, are considered.

In Figure 12a, it is represented the discharge along the channel at the time $t = 5$ s for the Riemann problem of Table 5, row 1. The comparison between exact (continuous line) and numerical solution (dashed line) demonstrates that the discharge is invariant through the geometric discontinuity in $x = 0$ m, as expected from the standing wave definition (see Sub-section 2.1). The same happens for the example of Table 5, row 2 (see Figure 12b), where the numerical discharge compares nicely with that of the exact solution. Actually, the numerical algorithm satisfies the condition of discharge invariance through the width discontinuity in all the numerical tests (not reported here) with the only important exception of the Riemann problems of Table 5, rows 6 and 8, whose exact solution is characterized by a hydraulic jump through the width discontinuity. The numerical discharge (dashed line) for these two tests is represented in Figure 12c and 12d, and compared with the corresponding exact solution (continuous line). The inspection of these two panels shows that the numerical solution exhibits a spike located in the first control volume to the right of the width expansion.

[Insert Figure 12 about here]

## 5. Discussion

In the present section, the solution multiplicity exhibited by the Riemann problem is discussed with reference to a similar phenomenon, namely the hydraulic hysteresis, observed in experimental hydraulics. Finally, the numerical results are discussed.

### 5.1 Multiple solutions and hydraulic hysteresis

It is clear that the existence of multiple solutions to the Riemann problem in variable-width channel is connected with the existence of multiple steady states for supercritical flows approaching a contraction (Cozzolino et al. 2018a). Actually, the Theorem 1 of Sub-section 3.4 states that the discontinuous-width Riemann problem may exhibit more than one solution for given initial conditions if $F_R \leq -K_{sp}(B_L/B_R)$ when $B_L < B_R$. This condition is satisfied when a supercritical flow approaches a contraction with energy greater than that strictly required to pass the geometric discontinuity (Yarnell 1934, see also Sub-section 2.1.1). The excess of energy, which can be dissipated by means of a standing hydraulic jump collocated into the contraction or by means of a backward moving shock, allows introducing three distinct steady flow conditions through the contraction in the one-dimensional theory (Akers and Bokhove 2008):

(T1) the supercritical flow smoothly passes through the contraction (standing wave SWe of Sub-section 2.1.2);

(T2) the supercritical flow is reverted into subcritical by means of a hydraulic jump collocated through the contraction, with reduction of the flow energy (standing wave SWf of Sub-section 2.1.2);

(T3) the supercritical flow is reverted into subcritical by means of a backward shock that moves upstream of the contraction; the subcritical flow, with reduced energy, passes through the contraction (standing wave SWd of Sub-section 2.1.2).

Akers and Bokhove (2008) have underlined that hydraulic jumps through contractions are unstable in the one-dimensional theory of channel flows when friction is absent and the channel is horizontal, thus implying the instability of all the Riemann solutions depending on the standing wave SWf. Following this observation, the reaches $J_2$-$\beta$-$\mathbf{u}_R^{0\#}$ on the curve III$_R$ and $\mathbf{u}_R^{\#0}$-$\mathbf{u}_R^{0\#}$ on the curve IV$_R$ should be canceled together with the corresponding Riemann solutions. Despite the apparent correctness of this conclusion, the standing wave definitions SWf and the corresponding Riemann solutions have been retained here not only for completeness but also to consider subtler effects emerging from laboratory experiments and theoretical work on two-dimensional Shallow water Equations. The experiments (Ackers and Bokhove 2008, Defina and Viero 2010) show that a supercritical flow approaching a channel contraction may assume one of the following three configurations:

(E1) the flow remains supercritical, but shock waves are originated from the edges of the contracting walls; these shock waves interact forming a distinctive two-dimensional shock pattern that is well studied in classic hydraulics literature (Ippen and Dawson 1951);

(E3) two shock waves originate from the edges of the contracting walls and join a transverse shock front, called Mach stem;

(E3) the flow through the contraction is subcritical, while a moving shock propagates upstream (chocked flow condition);

While the experimental condition E3 clearly coincides with the theoretical flow configuration T3 (standing wave SWd), the experimental conditions E1 and E2 somehow recall the theoretical flow configurations T1 (SWe) and T2 (SWf), respectively. Defina and Viero (2010) demonstrated that the E2 condition with downstream subcritical flow was stable if the channel bed was sloping and friction effects were taken into account, while the same condition with side supercritical flow was stable even on horizontal frictionless bed.

The coexistence of multiple steady states for supercritical flows approaching a contraction is called hydraulic hysteresis (Defina and Viero 2010, Viero and Defina 2017). In the early studies (Akers and Bokhove 2008, Defina and Viero 2010), it was assumed that this phenomenon depended on the upstream flow conditions and on the contraction characteristics, assuming that critical conditions were established at the downstream end of the channel contraction in the flow configurations E1 and E3 (classic hydraulic hysteresis). Viero and Defina (2017) have recently generalized the classic hydraulic hysteresis concept to the case that a subcritical downstream boundary condition may have a role in determining the flow characteristics through the channel contraction (generalized hydraulic hysteresis).

Notably, the Riemann solutions presented here are connected to both classic and generalized hydraulic hysteresis. For example, the Riemann problem of Table 5, row 10, exhibits critical conditions at the end of the contraction when the standing waves SWd (Figure 8a) and SWf (Figure 8b) are assumed through the geometric transition (classic hydraulic hysteresis). *Vice versa*, the Riemann problem of Table 5, row 11, exhibits subcritical conditions at the contraction end when SWd (Figure 9a) and SWf (Figure 9b) are assumed through the geometric transition (generalized hydraulic hysteresis). While the connection between multiple Riemann solutions and classic hydraulic hysteresis was already pointed out in Cozzolino et al. (2018a) and Varra et al. (2020), the connection with the generalized hydraulic hysteresis is underlined here for the first time.

## 5.2 Disambiguation of multiple solutions

When hydraulic hysteresis is established, it is possible to pass from one flow configuration to another by acting on the boundary conditions (Viero and Defina 2017) or by disturbing the flow (a Plexiglas paddle was used in Ackers and Bockhove 2008). In other words, the past history of the flow is important in determining the actual steady flow configuration through the contraction (Viero and Defina 2017). This is an interesting point where the theory of the Riemann problem with multiple solutions seems to depart from the theory of the hydraulic hysteresis. In fact, the Riemann problem of Eqs. (1)-(3) cancels the past history of the flow, while the boundary conditions cannot influence the solution because the channel considered in the Riemann problem has infinite length. This implies that only one of the alternative Riemann problem solutions can be chosen to enforce causality of the numerical models based on Riemann solution at cells' interfaces.

To disambiguate the Riemann problem solution, we observe that multiple solutions typically appear in hyperbolic systems of equations when non-conservative products take into account the interaction of the flow with solid boundaries. For example, the one-dimensional shallow water equations model with variable bed elevation (Han and Warnecke 2014), the one-dimensional Porous Shallow water Equations model (Cozzolino et al. 2018a), and the one-dimensional Shallow water Equations in variable-width channels (this paper), are all models that exhibit such a behaviour. The appearance of the geometric non-conservative products is due to the fact that the mathematical model at hand is obtained as a simplification of a mathematical model with higher dimensions. For example, the one-dimensional Shallow-water Equations model with variable bed elevation can be obtained from a three-dimensional model with solid boundary conditions, and the one-dimensional Shallow water Equations with variable width can be obtained from a two-dimensional Shallow water Equations model with horizontal bed and wall boundary conditions. This suggests the use of a higher-dimensions Riemann problem to pick up the physically relevant solution among the lower-dimensions alternatives (Warnecke and Andrianov 2004, Han et al. 2013).

**5.3 On the numerical solution**

The inspection of numerical results in Figure 10 and Figure 11 shows that the Finite Volume scheme by Cozzolino et al. (2018b) is able to capture the solution wave configuration structure in all the cases. Nonetheless, Figures 12c and 12d demonstrate that there is a residual computational error in the discharge through the geometric discontinuity when the solution exhibits a hydraulic jump through the channel expansion. It is evident that the numerical scheme does not correctly detect the presence of the hydraulic jump, inducing an unbalance between numerical fluxes and non-conservative products. This momentum imbalance is manifested as a localized discharge disturb. This error was not exhibited when the same algorithm was tested against a dam-break problem with hydraulic jump through the expansion (see Figure 9c in Cozzolino et al. 2018b). It follows that complete exact solutions of the Riemann problem, like that presented in this paper, are required to shed light on deficiencies of existing algorithms.

Interestingly, the examples of Figure 11d,e show that the algorithm by Cozzolino et al. (2018b) exhibits a tendency to pick the solution with supercritical flow through the geometric transition when multiple solutions are possible. This is congruent with the variables reconstruction at geometric discontinuities embedded into the algorithm, because this reconstruction assumes the standing wave SWe when the flow is supercritical. Nonetheless, there are clues from numerical experiments (Varra et al. 2020) that the two-dimensional Shallow water Equations exhibit chocked flow conditions when the one-dimensional problem admits multiple solutions. If this preliminary observation remains confirmed in future research, the algorithm should be modified in order to pick up this flow condition when multiple solutions are possible.

## 6. Conclusions

In the present paper, the complete solution to the Riemann problem of the one-dimensional variable-width Shallow water Equations has been presented. This solution is based on the assumption that the relationship between the states immediately to the left and to the right of the width discontinuity is a stationary weak solution of the Shallow water Equations. It is demonstrated that the solution to the Riemann problem always exists, and that it is unique in most case. However, there are certain initial conditions, characterized by supercritical flow impinging a contraction, where three solutions are possible. The appearance of multiple solutions to the discontinuous width Riemann problem is not surprising, and it is expected when non-conservative products are present in hyperbolic systems of equations.

In the case of the variable-width one-dimensional Shallow water Equations, the solution multiplicity is connected to a phenomenon, the hydraulic hysteresis, observed for supercritical flow in contracting channel. When hydraulic hysteresis is established in a channel, the steady state flow admits multiple different solutions, and it is possible to pass from one stable hydraulic condition to another by acting on the boundary conditions or by disturbing the flow. Of course, the Riemann problem is an initial value problem that cancels the past history of the flow, and boundary conditions are at infinite distance from the initial ones. This implies that only one of the Riemann solutions can be chosen. The problem of disambiguating the one-dimensional Riemann problem when multiple solutions are possible remains open, but the Authors' conjecture is that higher-dimensions mathematical problems can be used to pick up the physically relevant solution.

The analysis of a Finite Volume numerical scheme from the literature (Cozzolino et al. 2018b), which is based on variables' reconstruction at the finite volume interface assuming invariant energy and discharge, shows that the algorithm captures the solution with supercritical flow through the width discontinuity when multiple solutions are possible. If the supercritical solution is not the physical congruent one, the algorithm should be changed accordingly. Additionally, it has been found that it is very difficult to capture accurately those solutions that exhibit a hydraulic jump through the geometric discontinuity.

Interestingly, the one-dimensional variable-width Shallow water Equations are formally identical to the one-dimensional Porous Shallow water Equations, implying that the exact solutions and the numerical scheme discussed in the present paper are relevant for two-dimensional Porous Shallow water numerical models aiming at simulating urban flooding events. The exact solution presented here may be used not only as a benchmark, but it also supplies a guide for the construction of new algorithms, and it can be even embedded in an exact solver.

**Appendix A. Elementary waves of classic one-dimensional Shallow water Equations**

In the ($h$, $u$) plane, the curves of the direct shocks $S_i(\mathbf{u}_{ref})$ and the backward shocks $S_i^B(\mathbf{u}_{ref})$ contained into the $i$-th characteristic field of the classic one-dimensional Shallow water Equations are expressed by

(A.1) $S_i(\mathbf{u}_{ref})$, $S_i^B(\mathbf{u}_{ref})$: $u = u_{ref} \mp (h - h_{ref})\sqrt{0.5g(1/h + 1/h_{ref})}$,

where the minus sign is used for $i = 1$ while the plus sign is used for $i = 2$.

The curves of the direct rarefactions $R_i(\mathbf{u}_{ref})$ and the backward rarefactions $R_i^B(\mathbf{u}_{ref})$ contained into the $i$-th characteristic field are expressed by

(A.2) $R_i(\mathbf{u}_{ref})$, $R_i^B(\mathbf{u}_{ref})$: $u = u_{ref} \pm \left(2\sqrt{gh_{ref}} - 2\sqrt{gh}\right)$,

where the plus sign is used for $i = 1$ while the minus sign is used for $i = 2$.

**Appendix B. Froude limits and related curves**

For $B_0/B \leq 1$, the subcritical Froude limit is calculated as

(B.1) $K_{sb}(B_0/B) = (B_0/B)^{-1/2} \left[2\cos\left(\frac{5\pi}{3} - \frac{1}{3}\arctan\sqrt{\left(\frac{B_0}{B}\right)^{-1/2} - 1}\right)\right]^{\frac{3}{2}}$,

while the supercritical limit has the expression

(B.2) $K_{sp}(B_0/B) = (B_0/B)^{-1/2} \left[2\cos\left(\frac{\pi}{3} - \frac{1}{3}\arctan\sqrt{\left(\frac{B_0}{B}\right)^{-1/2} - 1}\right)\right]^{\frac{3}{2}}$.

The conjugate Froude numbers $K_{jump}(B_0/B)$ and $K_{sp}^{\#}(B_0/B)$ are defined as follows:

$$\begin{aligned}
K_{jump}(B_0/B) &= K_{sb}(B_0/B)\sqrt{8}\left(-1+\sqrt{1+8K_{sb}^2(B_0/B)}\right)^{-\frac{3}{2}} \\
K_{sp}^{\#}(B_0/B) &= K_{sp}(B_0/B)\sqrt{8}\left(-1+\sqrt{1+8K_{sp}^2(B_0/B)}\right)^{-\frac{3}{2}}
\end{aligned} \qquad (B.3)$$

**Figures**

Figure 1. Conceptual representation of the width discontinuity (monotonic smooth expansion).

Figure 2. Sketch of the admissible flow conditions through the width discontinuity. Flow through an expansion: standing wave SWa (a), SWb (b), SWc (c). Flow through a contraction: SWd (d), Swe I, SWf (f). The arrow represents the flow direction.

Figure 3. Example L-M curves for $B_L/B_R = 0.6$: curve $I_L$ for the state $\mathbf{u}_L$ in row 1 of Table 3 (a); curve $II_{L,a}$ for the state $\mathbf{u}_L$ in row 2 of Table 3 (b); curve $II_{L,b}$ for the state $\mathbf{u}_L$ in row 3 of Table 3 (c); curve $III_L$ for the state $\mathbf{u}_L$ in row 4 of Table 3 (d).

Figure 4. Example R-M curves for $B_L/B_R = 0.6$: curve $I_R$ for the state $\mathbf{u}_R$ in row 1 of Table 4 (a); curve $II_{R,a}$ for the state $\mathbf{u}_R$ in row 2 of Table 4 (b); curve $II_{R,b}$ for the state $\mathbf{u}_R$ in row 3 of Table 4 (c); curve $II_{R,c}$ for the state $\mathbf{u}_R$ in row 4 of Table 4 (d); curve $III_R$ for the state $\mathbf{u}_R$ in row 5 of Table 4 (e); curve $IV_R$ for the state $\mathbf{u}_R$ in row 6 of Table 4 (f).

Figure 5. Example Riemann problem 1 (initial conditions in Table 5, row 1): construction of the exact solution (a); flow-depth exact solution at time $t = 5$ s (b). Example Riemann problem 2 (Table 5, row 2): construction of the exact solution (c); flow-depth exact solution at time $t = 5$ s (d). Example Riemann problem 3 (Table 5, row 3): construction of the exact solution I; flow-depth exact solution at time $t = 5$ s (f). The black arrow depicts the direction of the flow through the geometric discontinuity.

Figure 6. Example Riemann problem 4 (initial conditions in Table 5, row 4): construction of the exact solution (a); flow-depth exact solution at time t = 5 s (b). Example Riemann problem 5 (Table 5, row 5): construction of the exact solution (c); flow-depth exact solution at time t = 5 s (d). Example Riemann problem 6 (Table 5, row 6): construction of the exact solution I; flow-depth exact solution at time t = 5 s (f). The black arrow depicts the direction of the flow through the geometric discontinuity.

Figure 7. Example Riemann problem 7 (initial conditions in Table 5, row 7): construction of the exact solution (a); flow-depth exact solution at time t = 5 s (b). Example Riemann problem 8

(Table 5, row 8): construction of the exact solution (c); flow-depth exact solution at time t = 5 s (d). Example Riemann problem 9 (Table 5, row 9): construction of the exact solution I; flow-depth exact solution at time t = 5 s (f). The black arrow depicts the direction of the flow through the geometric discontinuity.

Figure 8. Example Riemann problem 10 (initial conditions in Table 5, row 10). Construction of the first exact solution (a) and corresponding flow-depth exact solution at time t = 5 s (b). Construction of the second exact solution (c) and corresponding flow-depth exact solution at time t = 5 s (d). Construction of the third exact solution (e) and corresponding flow-depth exact solution at time t = 5 s (f). The black arrow depicts the direction of the flow through the geometric discontinuity.

Figure 9. Example Riemann problem 11 (initial conditions in Table 5, row 11). Construction of the first exact solution (a) and corresponding flow-depth exact solution at time t = 5 s (b). Construction of the second exact solution (c) and corresponding flow-depth exact solution at time t = 5 s (d). Construction of the third exact solution (e) and corresponding flow-depth exact solution at time t = 5 s (f). The black arrow depicts the direction of the flow through the geometric discontinuity.

Figure 10. Example Riemann problems from 1 to 6 (initial conditions in Table 5, rows from 1 to 6): Example Riemann problem 1 (a); example Riemann problem 2 (b); example Riemann problem 3 (c); example Riemann problem 4 (d); example Riemann problem 5 (e); example Riemann problem 6 (f). Exact (continuous black line) and numerical solution (dashed black line) for the flow depth at time t = 5 s.

Figure 11. Example Riemann problems from 7 to 11 (initial conditions in Table 5, rows from 7 to 11): Example Riemann problem 7 (a); example Riemann problem 8 (b); example Riemann problem 9 (c); example Riemann problem 10 (d); example Riemann problem 11 (e). Exact (continuous black line) and numerical solution (dashed black line) for the flow depth at time t = 5 s.

Figure 12. Example Riemann problems 1, 2, 6, and 8 (initial conditions in Table 5, rows 1, 2, 6, and 8): Example Riemann problem 1 (a); example Riemann problem 2 (b); example Riemann problem 6 (c); example Riemann problem 8 (d). Exact (continuous black line) and numerical solution (dashed black line) for the discharge at time t = 5 s.

**Tables**

Table 1. Solution configuration classes.

Table 2. Nomenclature of L-M and R-M curves.

Table 3. L-M curve examples.

Table 4. R-M curve examples.

Table 5. Initial conditions of the example Riemann problems.

Table 6. Intersection of the L-M and R-M curves for the example Riemann problems.

Table 7. Solution structure of the example Riemann problems.

Table 1. Solution configuration classes.

| Row | Code | Wave configuration sketch | Notes |
|---|---|---|---|
| 1 | - | $\mathbf{u}_L \xrightarrow{T_1} \mathbf{u}_{mid} \xrightarrow{T_2} \mathbf{u}_R$ | Classic Riemann problem ($B_L = B_R$) with intermediate state $\mathbf{u}_{mid}$. |
| 2 | - | $\mathbf{u}_L \xrightarrow{R_1} \mathbf{I}, \quad \mathbf{J} \xrightarrow{R_2} \mathbf{u}_R$ | Classic Riemann problem ($B_L = B_R$) with formation of dry bed state. |
| 3 | SC4+ | $\mathbf{u}_L \xrightarrow{R_1} \mathbf{u}_1 \xrightarrow{SWc} \mathbf{u}_2 \xrightarrow{T_1} \mathbf{u}_{mid} \xrightarrow{T_2} \mathbf{u}_R$ | $F_1 = 1, F_2 = K_{sp}(B_L/B_R)$ |
| 4 | SC3$_0$+ | $\mathbf{u}_L \xrightarrow{SWc} \mathbf{u}_2 \xrightarrow{T_1} \mathbf{u}_{mid} \xrightarrow{T_2} \mathbf{u}_R$ | $\mathbf{u}_1 = \mathbf{u}_L$ <br> $F_1 > 1, F_2 > K_{sp}(B_L/B_R)$ |
| 5 | SC3+ | $\mathbf{u}_L \xrightarrow{T_1} \mathbf{u}_1 \xrightarrow{SW} \mathbf{u}_2 \xrightarrow{T_2} \mathbf{u}_R$ | 1) SWa, $0 \leq F_1 < 1, 0 \leq F_2 < K_{sb}(B_L/B_R)$ <br> 2) SWb, $F_1 = 1, K_{sb}(B_L/B_R) < F_2 < K_{sp}^{\#}(B_L/B_R)$ |
| 6 | SC2$_0$+ | $\mathbf{u}_L \xrightarrow{SWb} \mathbf{u}_2 \xrightarrow{T_2} \mathbf{u}_R$ | $\mathbf{u}_1 = \mathbf{u}_L$ <br> $F_1 > 1, 0 < F_2 < 1$ |
| 7 | SC4- | $\mathbf{u}_L \xrightarrow{T_1} \mathbf{u}_{mid} \xrightarrow{R_2} \mathbf{u}_1 \xrightarrow{SWd} \mathbf{u}_2 \xrightarrow{T_2} \mathbf{u}_R$ | $F_1 = -1, F_2 = -K_{sb}(B_L/B_R)$ |
| 8 | SC3$_0$- | $\mathbf{u}_L \xrightarrow{T_1} \mathbf{u}_{mid} \xrightarrow{T_2} \mathbf{u}_1 \xrightarrow{SW} \mathbf{u}_R$ | $\mathbf{u}_2 = \mathbf{u}_R$ <br> 1) SWe, $F_1 < -1, F_2 < -K_{sp}(B_L/B_R)$ <br> 2) SWf, $F_1 = -1, F_2 < -K_{sp}(B_L/B_R), T_2 = R_2$ |
| 9 | SC3- | $\mathbf{u}_L \xrightarrow{T_1} \mathbf{u}_1 \xrightarrow{SWd} \mathbf{u}_2 \xrightarrow{T_2} \mathbf{u}_R$ | $-1 \leq F_1 < 0, -K_{sb}(B_L/B_R) \leq F_2 < 0$ |
| 10 | SC2$_0$- | $\mathbf{u}_L \xrightarrow{T_1} \mathbf{u}_1 \xrightarrow{SWf} \mathbf{u}_R$ | $\mathbf{u}_2 = \mathbf{u}_R$ <br> $-1 < F_1 < 0, F_2 < -K_{sp}(B_L/B_R)$ |

Table 2. Nomenclature of L-M and R-M curves.

| Row | L-M curve | $F_L$ limits |
|---|---|---|
| 1 | $I_L$ | $F_L < -2$ |
| 2 | $II_L$ | $F_L \in [-2, 1]$ |
| 3 | $III_L$ | $F_L > 1$ |
| Row | R-M curve | $F_R$ limits |
| 4 | $I_R$ | $F_R > 2$ |
| 5 | $II_R$ | $F_R \in \,]-K_{sp}(B_L/B_R), 2]$ |
| 6 | $III_R$ | $F_R \in \,]-K_{jump}(B_L/B_R), -K_{sp}(B_L/B_R)]$ |
| 7 | $IV_R$ | $F_R \leq -K_{jump}(B_L/B_R)$ |

Table 3. L-M curve examples.

| Row | L-M curve | $h_L$ (m) | $u_L$ (m/s) | $F_L$ | Figure |
|---|---|---|---|---|---|
| 1 | $I_L$ | 1.00 | -8.00 | -2.55 | 3a |
| 2 | $II_{L,a}$ | 1.00 | -2.00 | -0.64 | 3b |
| 3 | $II_{L,b}$ | 1.00 | 2.00 | 0.64 | 3c |
| 4 | $III_L$ | 1.00 | 5.00 | 1.60 | 3d |

Table 4. R-M curve examples.

| Row | R-M curve | $h_R$ (m) | $u_R$ (m/s) | $F_R$ | Figure |
|---|---|---|---|---|---|
| 1 | $I_R$ | 1.00 | 8.00 | 2.55 | 4a |
| 2 | $II_{R,a}$ | 1.00 | 2.00 | 0.64 | 4b |
| 3 | $II_{R,b}$ | 1.00 | -0.50 | -0.16 | 4c |
| 4 | $II_{R,c}$ | 1.00 | -5.00 | -1.60 | 4d |
| 5 | $III_R$ | 1.00 | -9.40 | -3.00 | 4e |
| 6 | $IV_R$ | 1.00 | -13.0 | -4.15 | 4f |

Table 5. Initial conditions of the example Riemann problems.

| Row | Example | $h_L$ (m) | $u_L$ (m/s) | $h_R$ (m) | $u_R$ (m/s) | L-M curve | R-M curve | Figure |
|---|---|---|---|---|---|---|---|---|
| 1 | 1 | 1.00 | -8.00 | 1.00 | 2.00 | $I_L$ | $II_{R,a}$ | 5a |
| 2 | 2 | 1.00 | -2.00 | 1.00 | -0.50 | $II_{L,a}$ | $II_{R,b}$ | 5c |
| 3 | 3 | 1.00 | 2.00 | 1.00 | 2.00 | $II_{L,b}$ | $II_{R,a}$ | 5e |
| 4 | 4 | 1.00 | 2.00 | 1.00 | -0.50 | $II_{L,b}$ | $II_{R,b}$ | 6a |
| 5 | 5 | 1.00 | 2.00 | 1.00 | -5.00 | $II_{L,b}$ | $II_{R,c}$ | 6c |
| 6 | 6 | 1.00 | 5.00 | 1.00 | -0.50 | $III_L$ | $II_{R,b}$ | 6e |
| 7 | 7 | 1.00 | 5.00 | 1.00 | 2.00 | $III_L$ | $II_{R,a}$ | 7a |
| 8 | 8 | 1.00 | 2.00 | 1.00 | 1.50 | $II_{L,b}$ | $II_{R,a}$ | 7c |
| 9 | 9 | 0.30 | -10.00 | 1.00 | 2.00 | $I_L$ | $II_{R,a}$ | 7e |
| 10 | 10 | 1.00 | -2.00 | 1.00 | -9.40 | $II_{L,a}$ | $III_R$ | 8a,c,e |
| 11 | 11 | 1.00 | 7.00 | 1.00 | -13.00 | $III_L$ | $IV_R$ | 9a,c,e |

Table 6. Intersection of the L-M and R-M curves for the example Riemann problems.

| Row | Example | $h_M$ (m) | $h_M$ (m/s) | Figure |
|---|---|---|---|---|
| 1 | 1 | 0.051 | -3.15 | 5a |
| 2 | 2 | 0.83 | -1.45 | 5c |
| 3 | 3 | 0.86 | 1.55 | 5e |
| 4 | 4 | 1.37 | 0.59 | 6a |
| 5 | 5 | 2.42 | -1.74 | 6c |
| 6 | 6 | 1.79 | 1.68 | 6e |
| 7 | 7 | 1.32 | 2.93 | 7a |
| 8 | 8 | 0.95 | 1.34 | 7c |
| 9 | 9 | - | - | 7e |
| 10 | 10 | 2.45 | -5.81 | 8a |
| 11 |  | 2.55 | -6.06 | 8c |
| 12 |  | 2.76 | -6.55 | 8e |
| 13 | 11 | 5.26 | -3.30 | 9a |
| 14 |  | 5.36 | -3.91 | 9c |
| 15 |  | 5.83 | -4.60 | 9e |

Table 7. Solution structure of the example Riemann problems.

| Row | Example | Solution structure | Figure |
|---|---|---|---|
| 1 | 1 | $\mathbf{u}_L \xrightarrow{R_1} \mathbf{u}_{mid} \xrightarrow{R_2} \mathbf{u}_1 \xrightarrow{SWd} \mathbf{u}_2 \xrightarrow{R_2} \mathbf{u}_R$ | 5b |
| 2 | 2 | $\mathbf{u}_L \xrightarrow{R_1} \mathbf{u}_1 \xrightarrow{SWd} \mathbf{u}_2 \xrightarrow{R_2} \mathbf{u}_R$ | 5d |
| 3 | 3 | $\mathbf{u}_L \xrightarrow{R_1} \mathbf{u}_1 \xrightarrow{SWc} \mathbf{u}_2 \xrightarrow{S_1} \mathbf{u}_{mid} \xrightarrow{R_2} \mathbf{u}_R$ | 5f |
| 4 | 4 | $\mathbf{u}_L \xrightarrow{S_1} \mathbf{u}_1 \xrightarrow{SWa} \mathbf{u}_2 \xrightarrow{S_2} \mathbf{u}_R$ | 6b |
| 5 | 5 | $\mathbf{u}_L \xrightarrow{S_1} \mathbf{u}_1 \xrightarrow{SWd} \mathbf{u}_2 \xrightarrow{S_2} \mathbf{u}_R$ | 6d |
| 6 | 6 | $\mathbf{u}_L \xrightarrow{SWc} \mathbf{u}_2 \xrightarrow{S_2} \mathbf{u}_R$ | 6f |
| 7 | 7 | $\mathbf{u}_L \xrightarrow{SWc} \mathbf{u}_2 \xrightarrow{S_1} \mathbf{u}_{mid} \xrightarrow{S_2} \mathbf{u}_R$ | 7b |
| 8 | 8 | $\mathbf{u}_L \xrightarrow{R_1} \mathbf{u}_1 \xrightarrow{SWb} \mathbf{u}_2 \xrightarrow{R_2} \mathbf{u}_R$ | 7d |
| 9 | 9 | $\mathbf{u}_L \xrightarrow{R_1} \mathbf{I},\ \mathbf{J} \xrightarrow{R_2} \mathbf{u}_1 \xrightarrow{SWd} \mathbf{u}_2 \xrightarrow{R_2} \mathbf{u}_R$ | 7f |
| 10 | 10 | $\mathbf{u}_L \xrightarrow{S_1} \mathbf{u}_{mid} \xrightarrow{R_2} \mathbf{u}_1 \xrightarrow{SWd} \mathbf{u}_2 \xrightarrow{S_2} \mathbf{u}_R$ | 8b |
| 11 | | $\mathbf{u}_L \xrightarrow{S_1} \mathbf{u}_{mid} \xrightarrow{R_2} \mathbf{u}_1 \xrightarrow{SWf} \mathbf{u}_R$ | 8d |
| 12 | | $\mathbf{u}_L \xrightarrow{S_1} \mathbf{u}_{mid} \xrightarrow{S_2} \mathbf{u}_1 \xrightarrow{SWe} \mathbf{u}_R$ | 8f |
| 13 | 11 | $\mathbf{u}_L \xrightarrow{S_1} \mathbf{u}_{mid} \xrightarrow{S_2} \mathbf{u}_1 \xrightarrow{SWd} \mathbf{u}_R$ | 9b |
| 14 | | $\mathbf{u}_L \xrightarrow{S_1} \mathbf{u}_1 \xrightarrow{SWf} \mathbf{u}_R$ | 9d |
| 15 | | $\mathbf{u}_L \xrightarrow{S_1} \mathbf{u}_{mid} \xrightarrow{S_2} \mathbf{u}_1 \xrightarrow{SWe} \mathbf{u}_R$ | 9f |

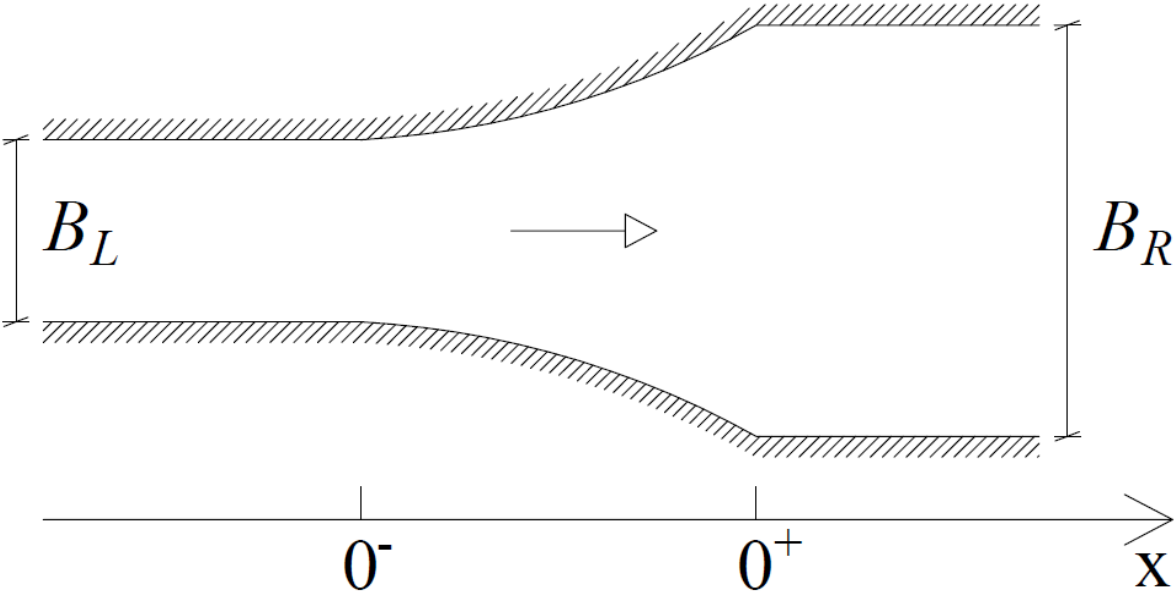

Figure 1. Conceptual representation of the width discontinuity (monotonic smooth expansion).

Figure 2. Sketch of the admissible flow conditions through the width discontinuity. Flow through an expansion: standing wave SWa (a), SWb (b), SWc (c). Flow through a contraction: SWd (d), Swe I, SWf (f). The arrow represents the flow direction.

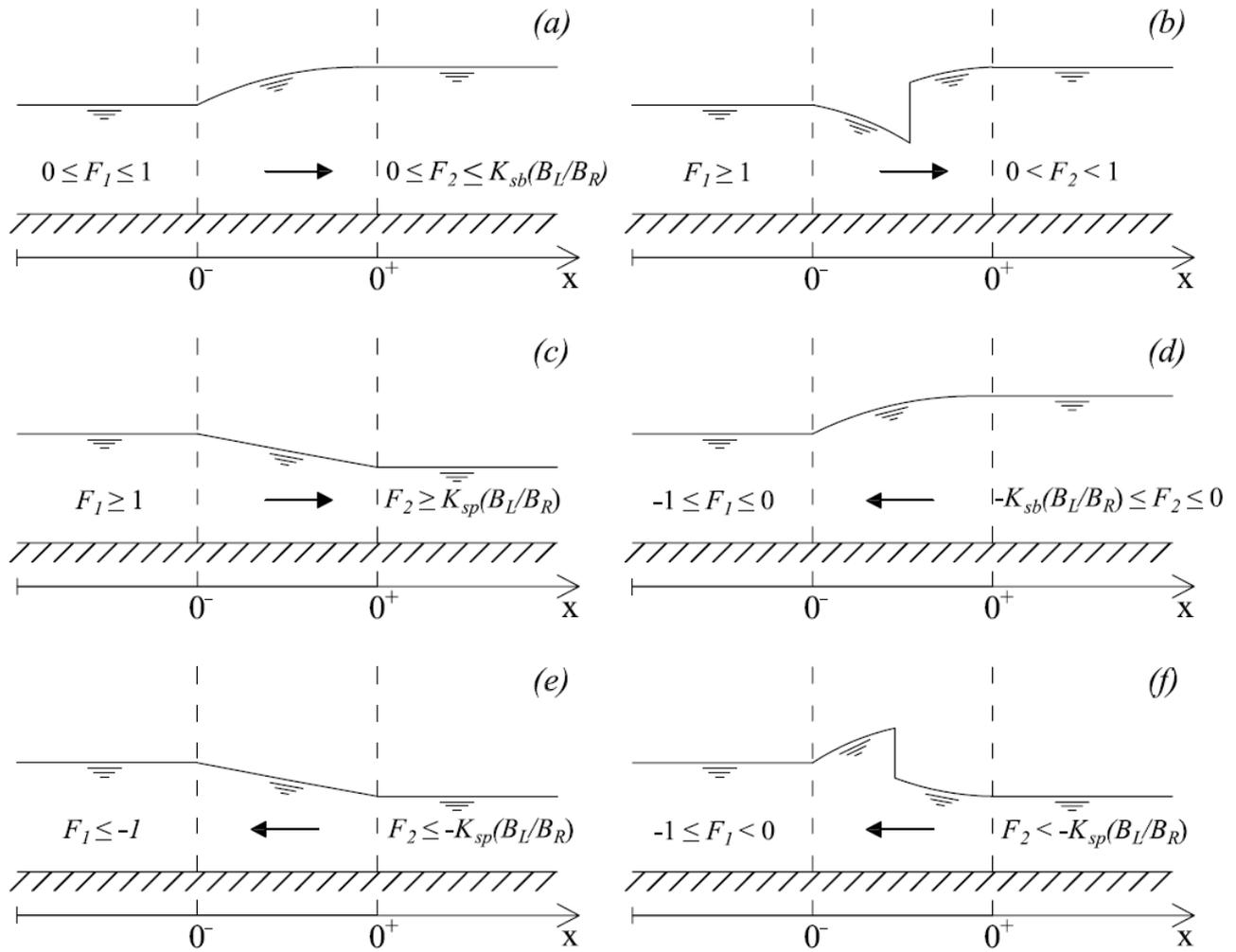

Figure 3. Example L-M curves for $B_L/B_R = 0.6$: curve $I_L$ for the state $\mathbf{u}_L$ in row 1 of Table 3 (a); curve $II_{L,a}$ for the state $\mathbf{u}_L$ in row 2 of Table 3 (b); curve $II_{L,b}$ for the state $\mathbf{u}_L$ in row 3 of Table 3 (c); curve $III_L$ for the state $\mathbf{u}_L$ in row 4 of Table 3 (d).

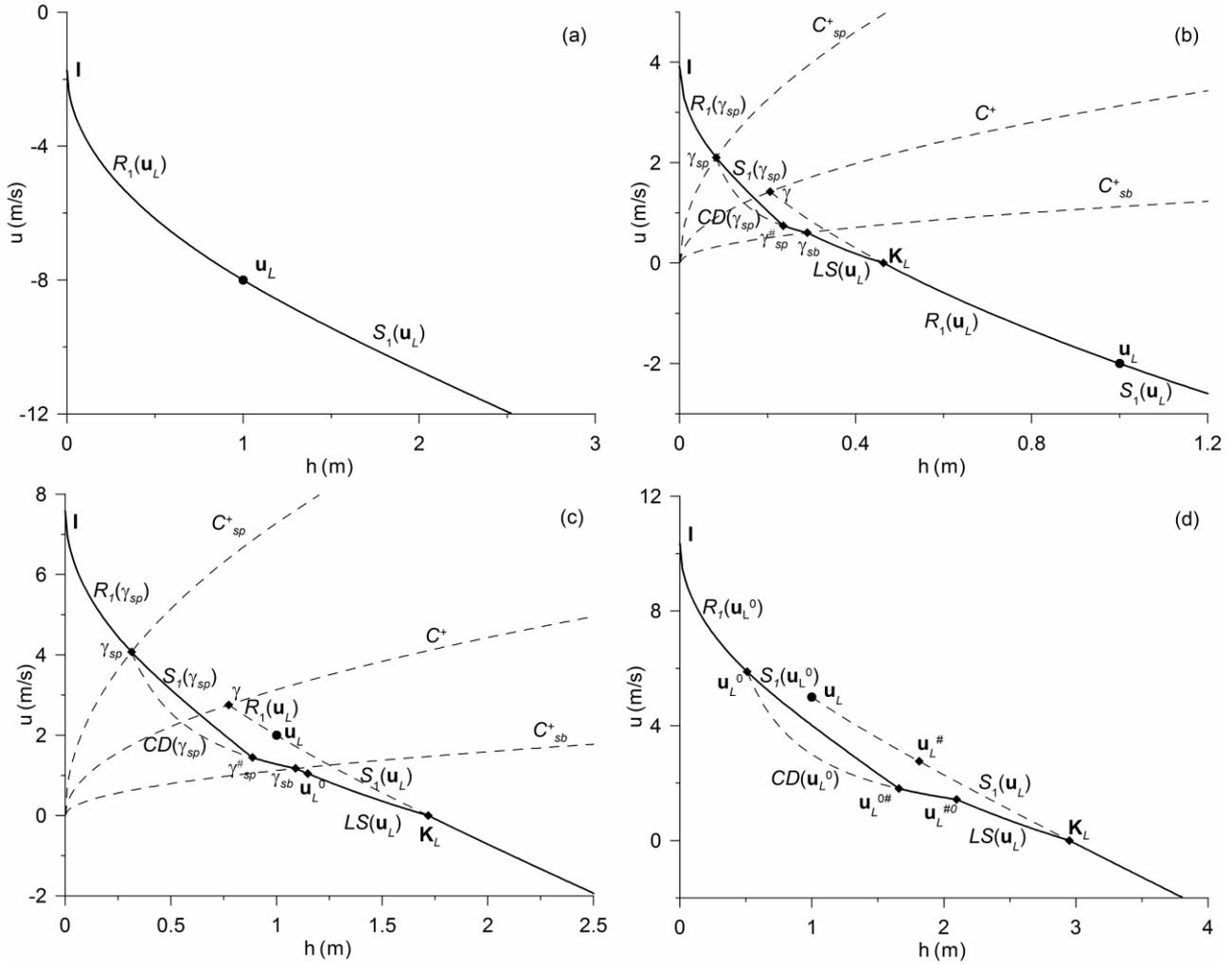

Figure 4. Example R-M curves for $B_L/B_R = 0.6$: curve $I_R$ for the state $\mathbf{u}_R$ in row 1 of Table 4 (a); curve $II_{R,a}$ for the state $\mathbf{u}_R$ in row 2 of Table 4 (b); curve $II_{R,b}$ for the state $\mathbf{u}_R$ in row 3 of Table 4 (c); curve $II_{R,c}$ for the state $\mathbf{u}_R$ in row 4 of Table 4 (d); curve $III_R$ for the state $\mathbf{u}_R$ in row 5 of Table 4 (e); curve $IV_R$ for the state $\mathbf{u}_R$ in row 6 of Table 4 (f).

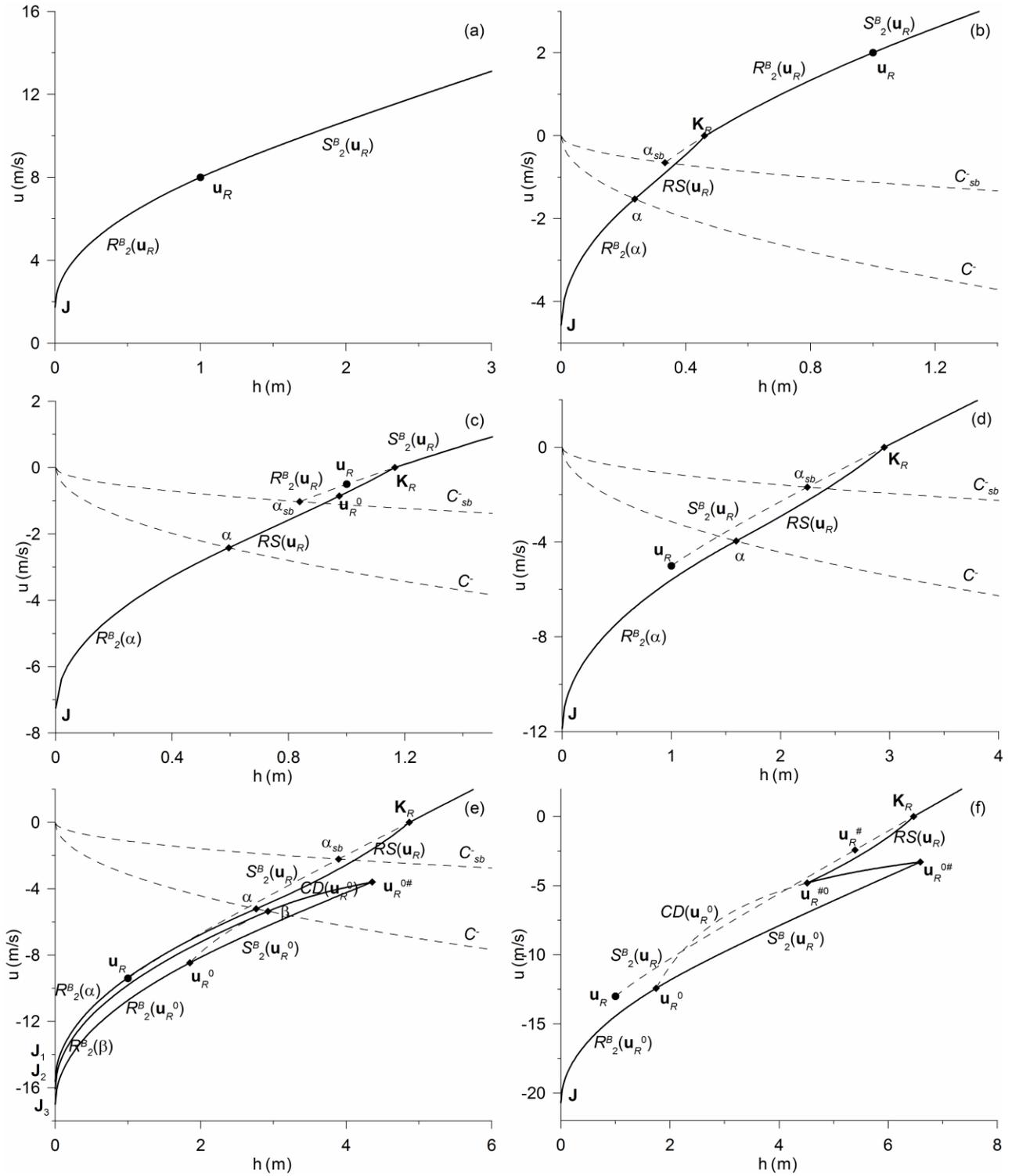

Figure 5. Example Riemann problem 1 (initial conditions in Table 5, row 1): construction of the exact solution (a); flow-depth exact solution at time $t = 5$ s (b). Example Riemann problem 2 (Table 5, row 2): construction of the exact solution (c); flow-depth exact solution at time $t = 5$ s (d). Example Riemann problem 3 (Table 5, row 3): construction of the exact solution (e); flow-depth exact solution at time $t = 5$ s (f). The black arrow depicts the direction of the flow through the geometric discontinuity.

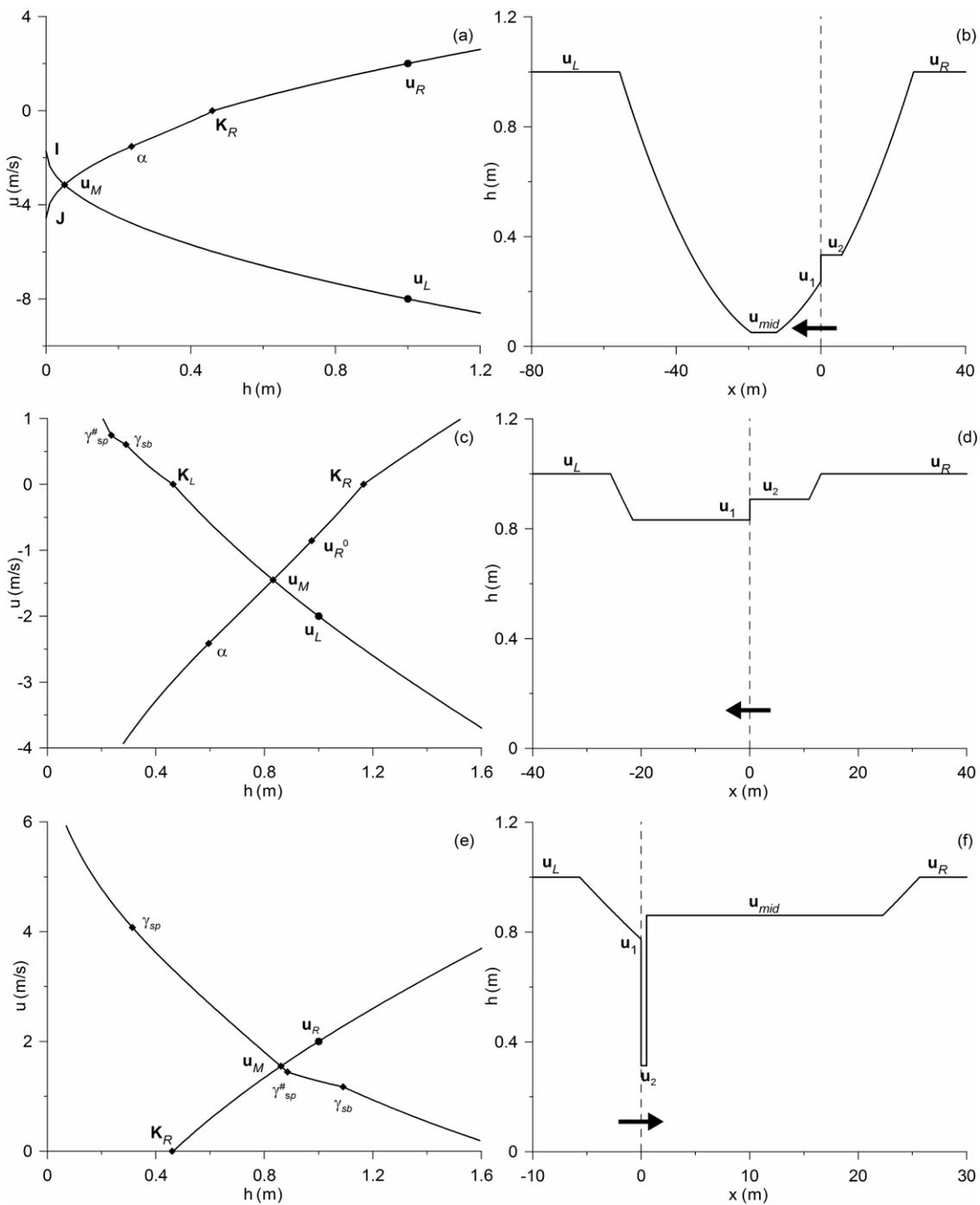

Figure 6. Example Riemann problem 4 (initial conditions in Table 5, row 4): construction of the exact solution (a); flow-depth exact solution at time $t = 5$ s (b). Example Riemann problem 5 (Table 5, row 5): construction of the exact solution (c); flow-depth exact solution at time $t = 5$ s (d). Example Riemann problem 6 (Table 5, row 6): construction of the exact solution (e); flow-depth exact solution at time $t = 5$ s (f). The black arrow depicts the direction of the flow through the geometric discontinuity.

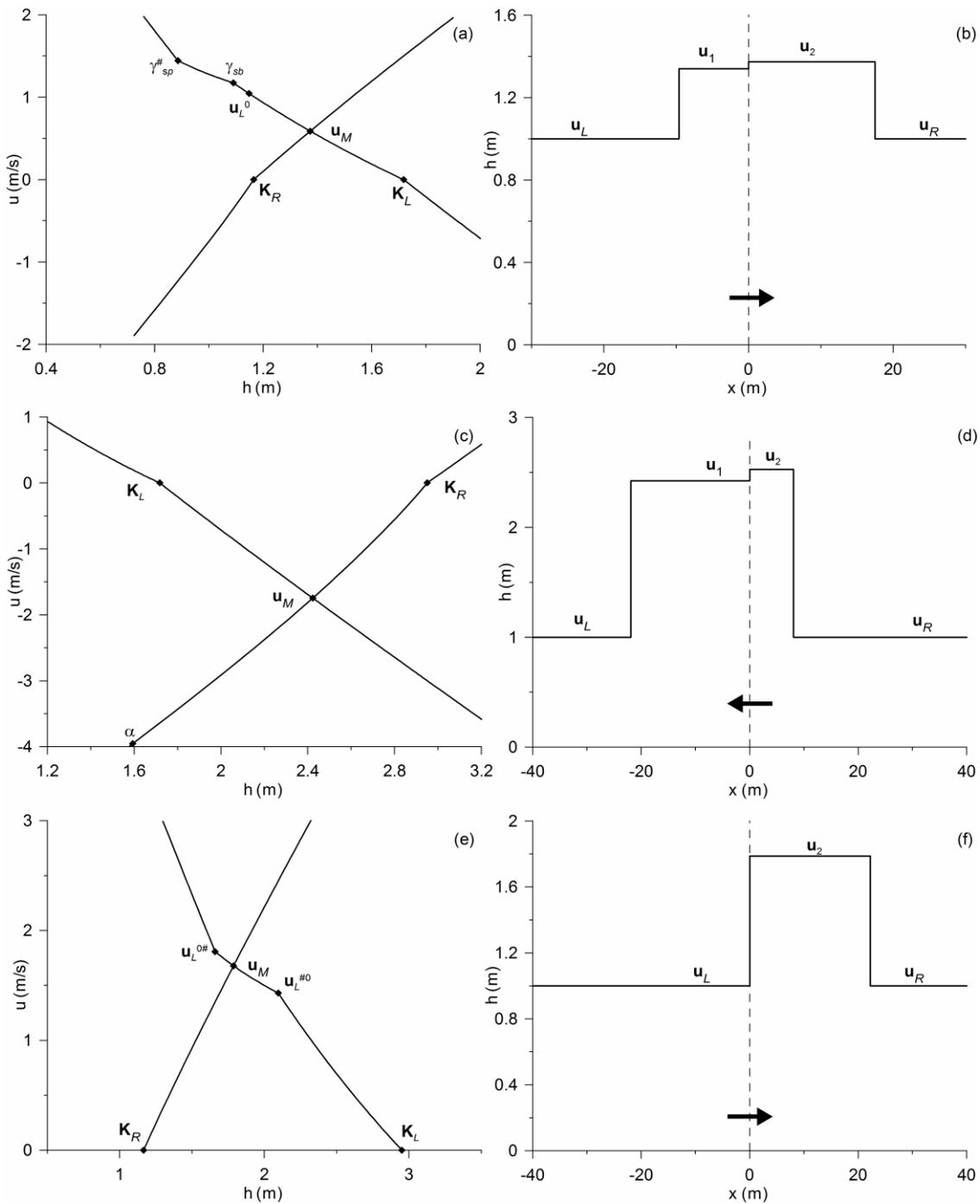

Figure 7. Example Riemann problem 7 (initial conditions in Table 5, row 7): construction of the exact solution (a); flow-depth exact solution at time $t = 5$ s (b). Example Riemann problem 8 (Table 5, row 8): construction of the exact solution (c); flow-depth exact solution at time $t = 5$ s (d). Example Riemann problem 9 (Table 5, row 9): construction of the exact solution (e); flow-depth exact solution at time $t = 5$ s (f). The black arrow depicts the direction of the flow through the geometric discontinuity.

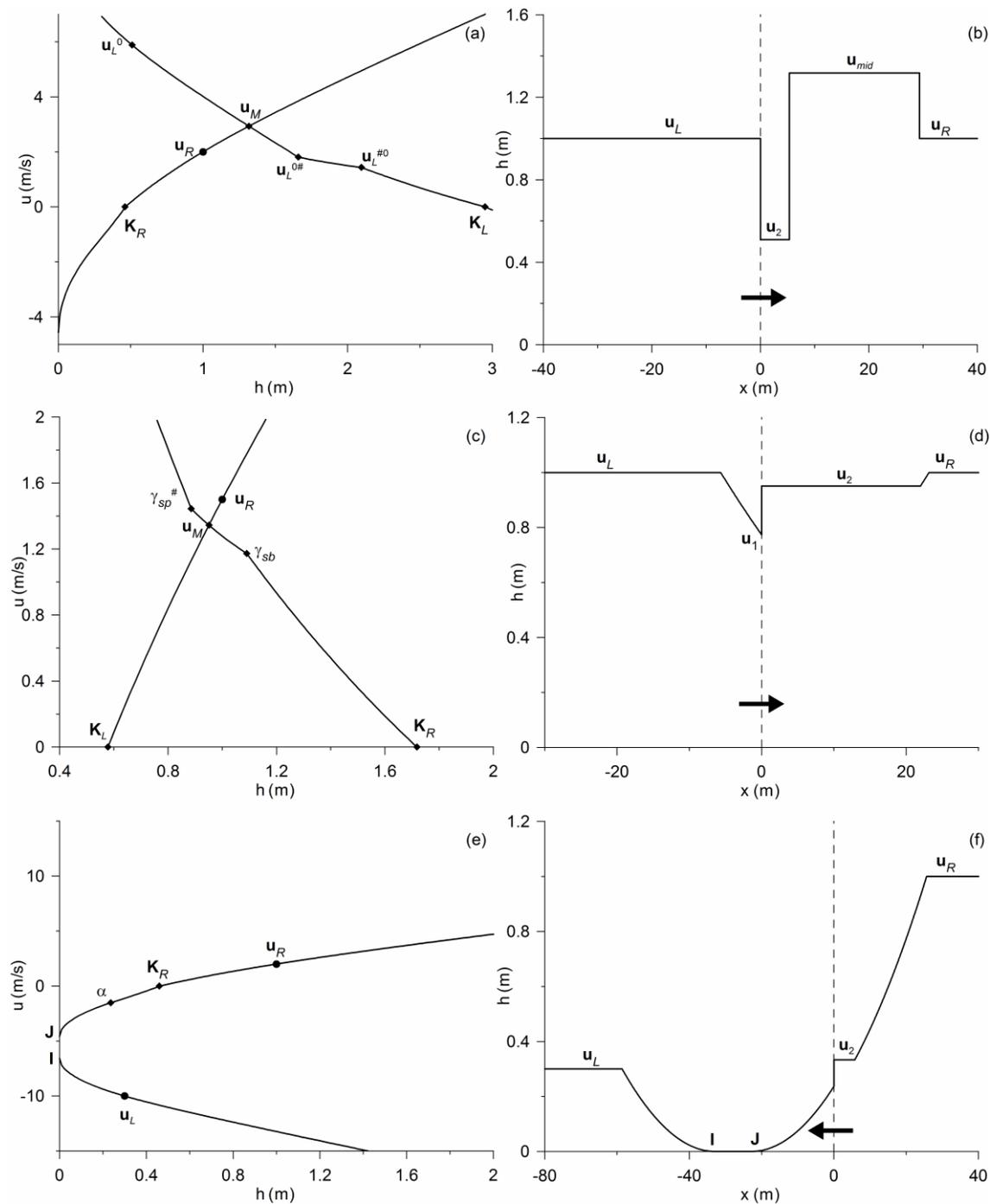

Figure 8. Example Riemann problem 10 (initial conditions in Table 5, row 10). Construction of the first exact solution (a) and corresponding flow-depth exact solution at time $t = 5$ s (b). Construction of the second exact solution (c) and corresponding flow-depth exact solution at time $t = 5$ s (d). Construction of the third exact solution (e) and corresponding flow-depth exact solution at time $t = 5$ s (f). The black arrow depicts the direction of the flow through the geometric discontinuity.

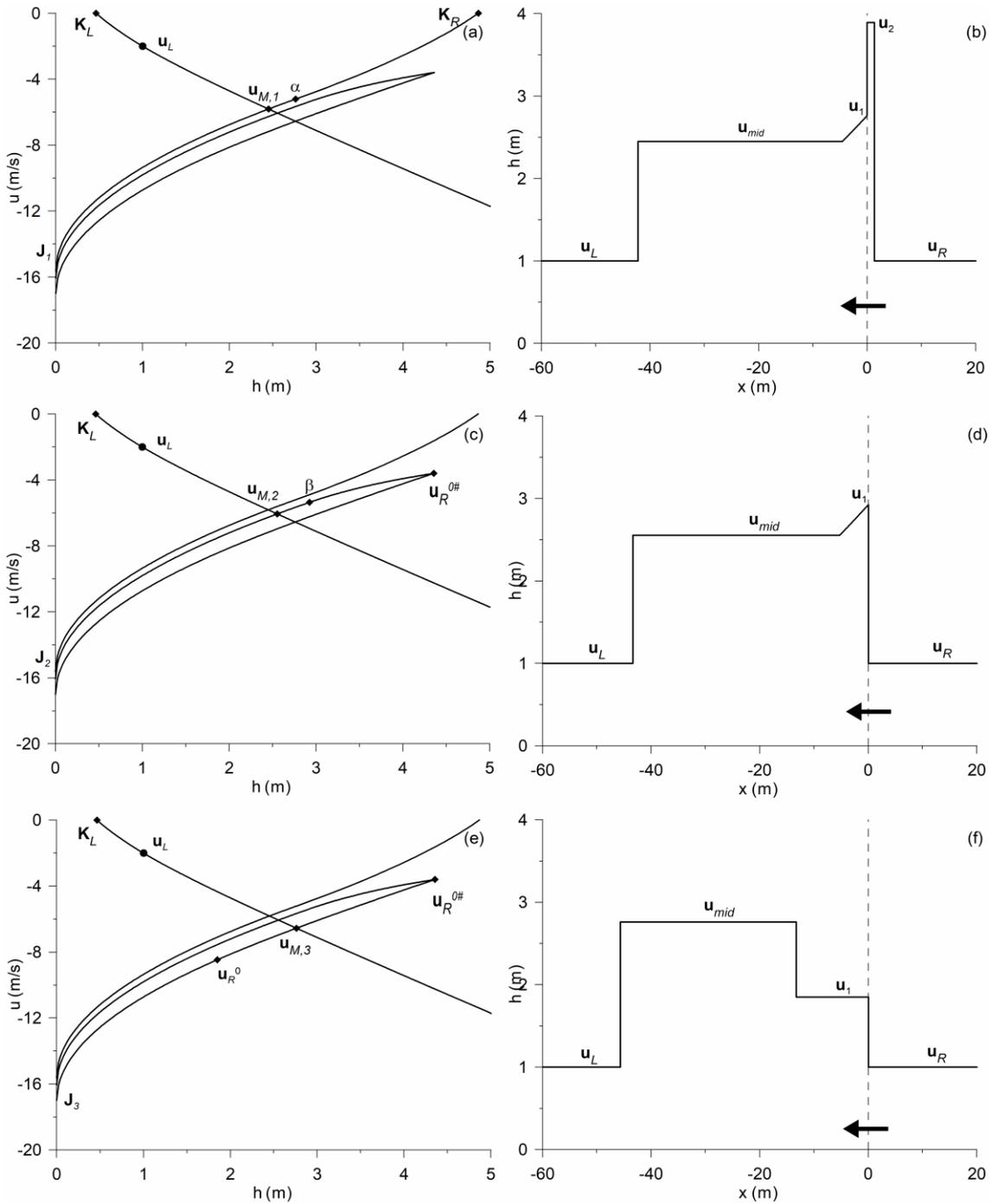

Figure 9. Example Riemann problem 11 (initial conditions in Table 5, row 11). Construction of the first exact solution (a) and corresponding flow-depth exact solution at time t = 5 s (b). Construction of the second exact solution (c) and corresponding flow-depth exact solution at time t = 5 s (d). Construction of the third exact solution (e) and corresponding flow-depth exact solution at time t = 5 s (f). The black arrow depicts the direction of the flow through the geometric discontinuity.

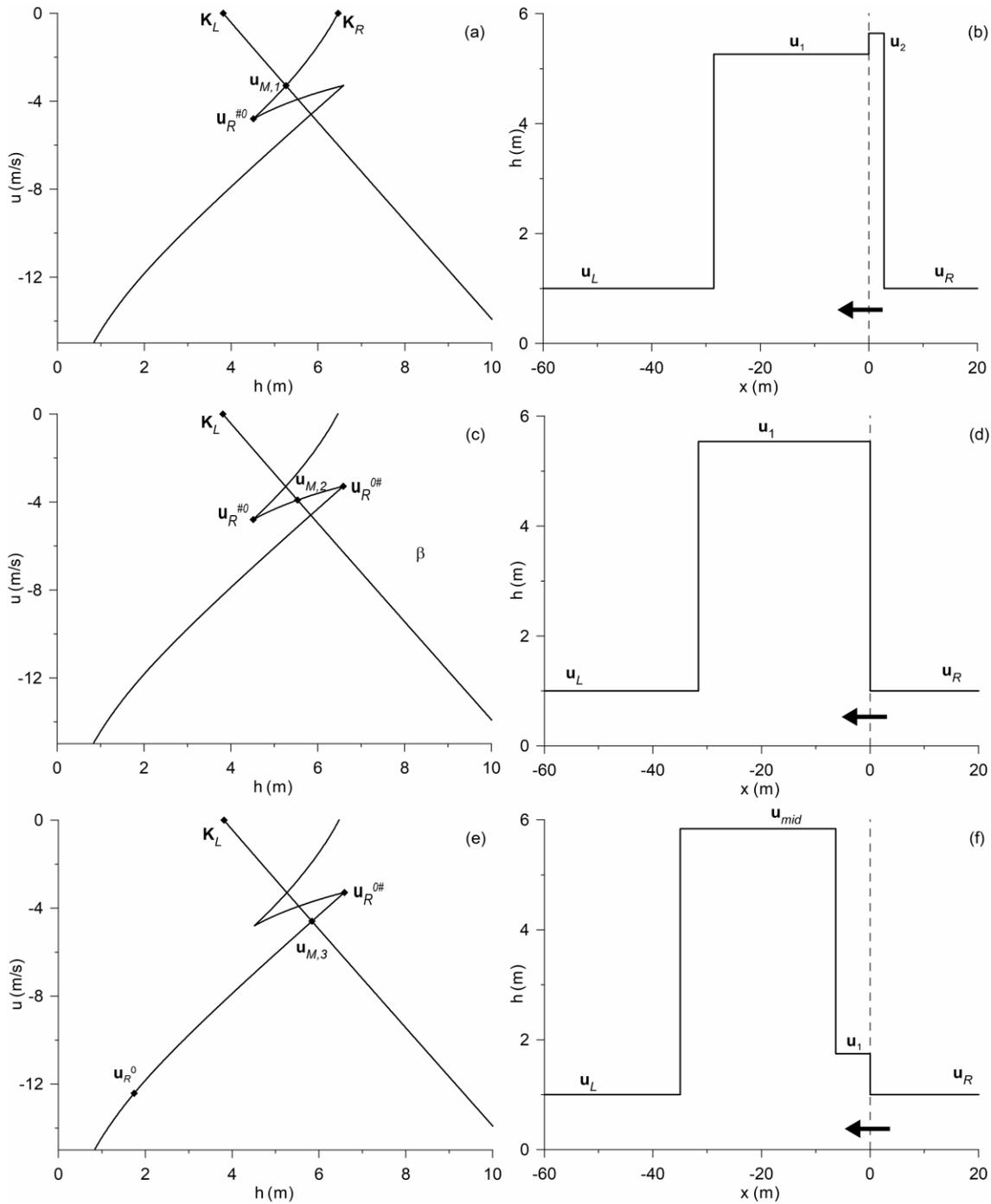

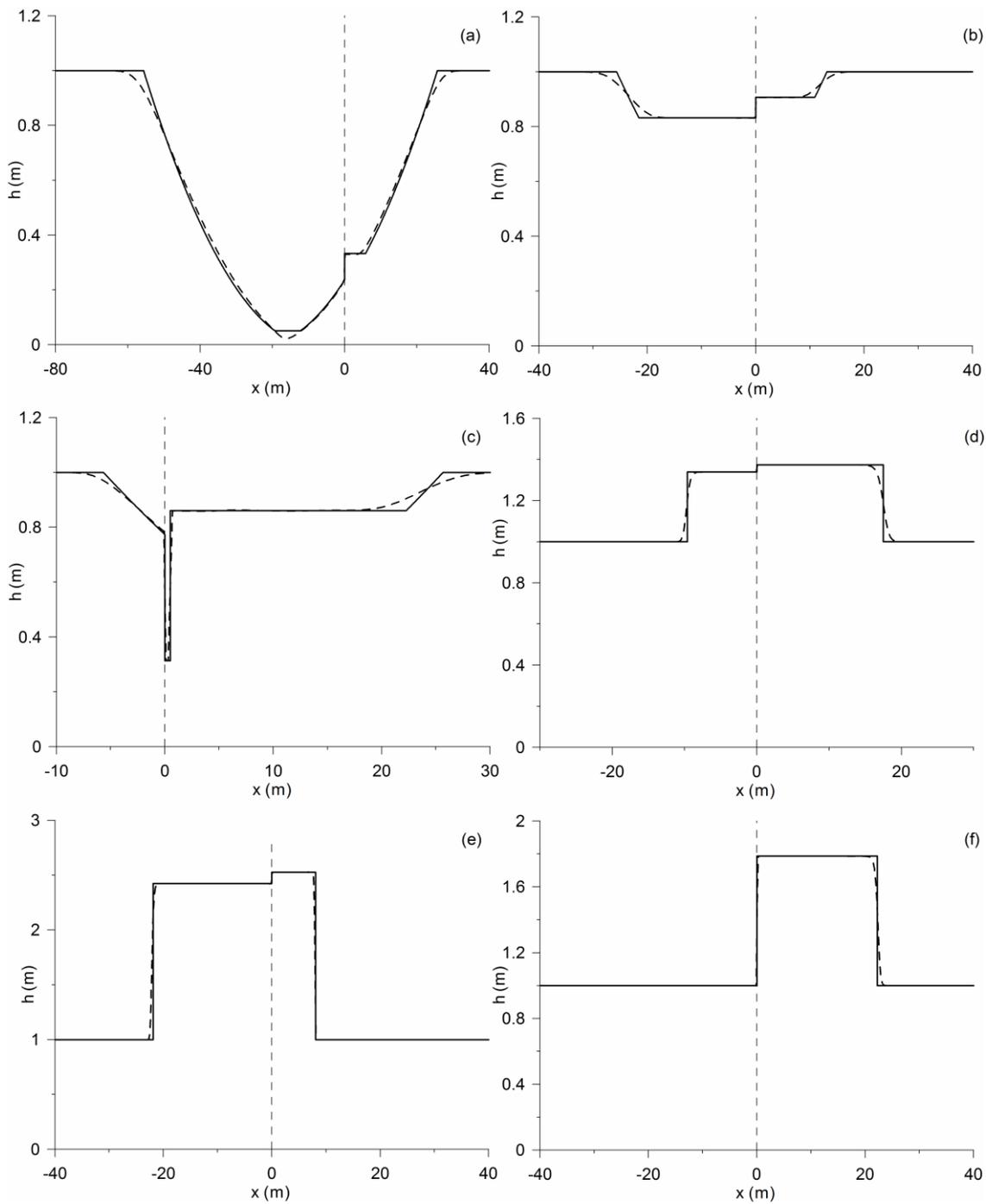

Figure 10. Example Riemann problems from 1 to 6 (initial conditions in Table 5, rows from 1 to 6): Example Riemann problem 1 (a); example Riemann problem 2 (b); example Riemann problem 3 (c); example Riemann problem 4 (d); example Riemann problem 5 (e); example Riemann problem 6 (f). Exact (continuous black line) and numerical solution (dashed black line) for the flow depth at time t = 5 s.

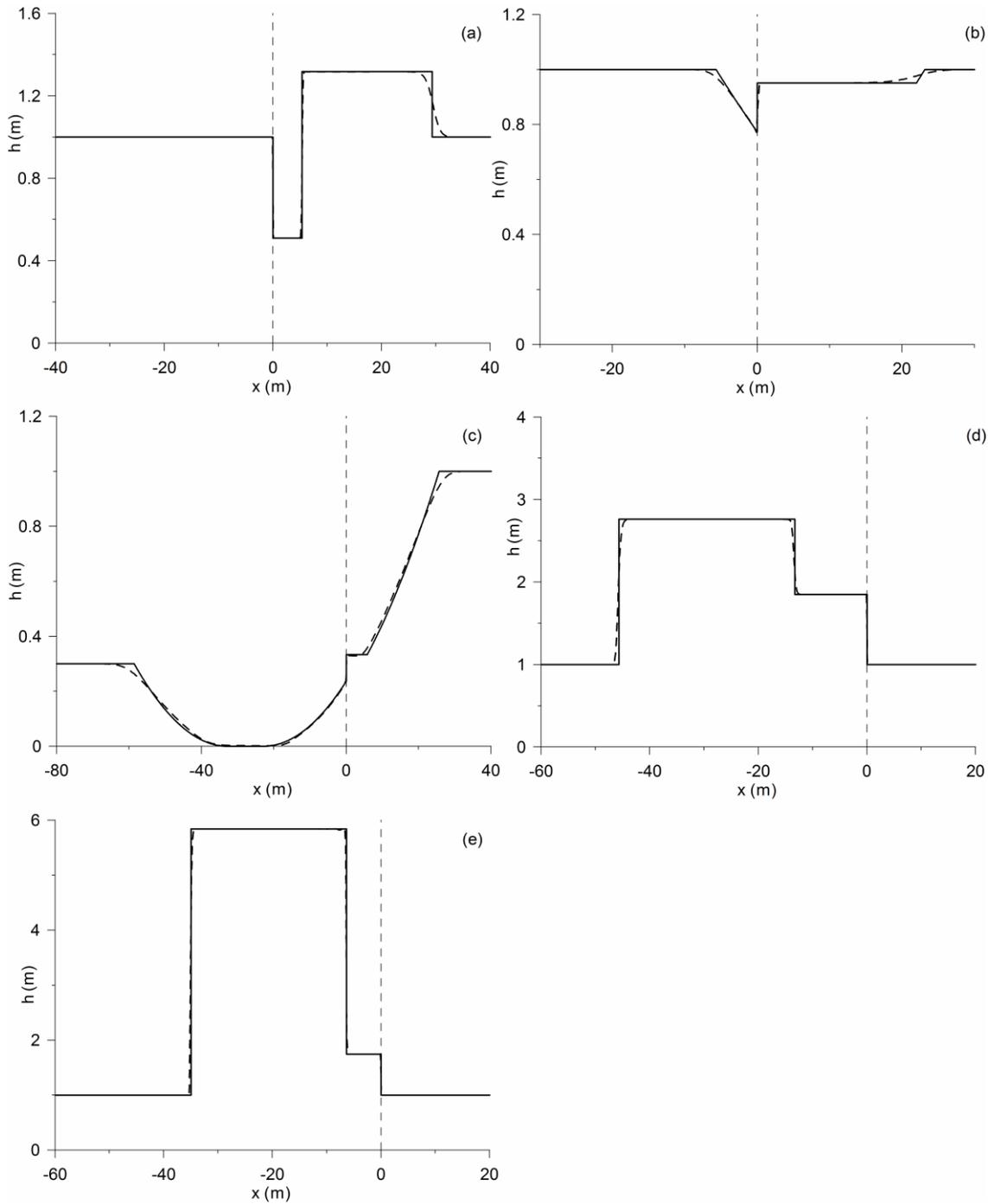

Figure 11. Example Riemann problems from 7 to 11 (initial conditions in Table 5, rows from 7 to 11): Example Riemann problem 7 (a); example Riemann problem 8 (b); example Riemann problem 9 (c); example Riemann problem 10 (d); example Riemann problem 11 (e). Exact (continuous black line) and numerical solution (dashed black line) for the flow depth at time t = 5 s.

Figure 12. Example Riemann problems 1, 2, 6, and 8 (initial conditions in Table 5, rows 1, 2, 6, and 8): Example Riemann problem 1 (a); example Riemann problem 2 (b); example Riemann problem 6 (c); example Riemann problem 8 (d). Exact (continuous black line) and numerical solution (dashed black line) for the discharge at time t = 5 s.

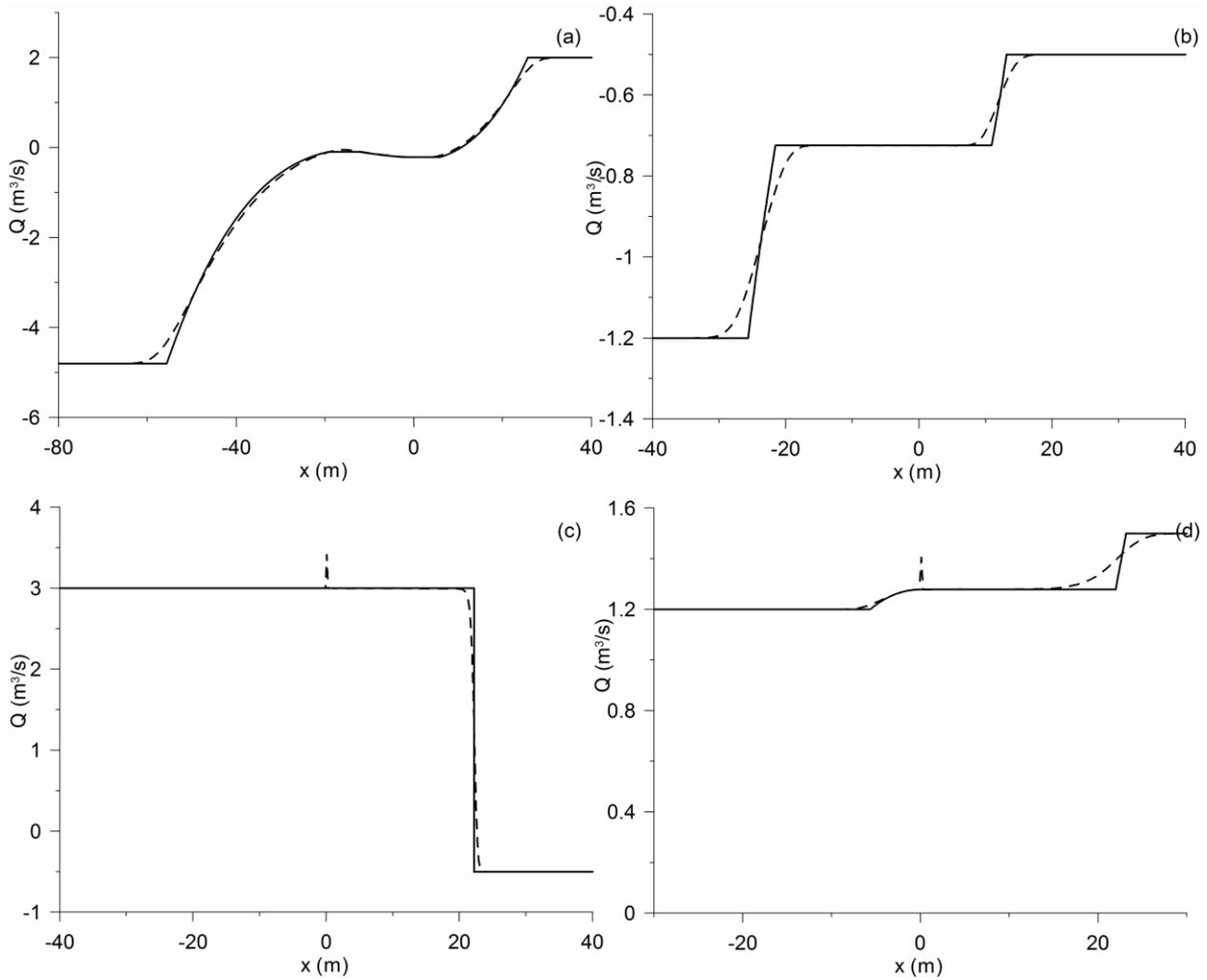